# Asymmetric Energy Landscapes Control Diffusion in Glasses

Ajay Annamareddy,[1,*] Bu Wang,[2,1] Paul M. Voyles,[1] Izabela Szlufarska,[1] and Dane Morgan[1,*]

[1]*Department of Materials Science and Engineering, University of Wisconsin–Madison, Madison, Wisconsin 53706, USA*

[2]*Department of Civil and Environmental Engineering, University of Wisconsin–Madison, Madison, Wisconsin 53706, USA*

[*]Authors to whom correspondence should be addressed: ajeykrishna@gmail.com and ddmorgan@wisc.edu

While diffusion in crystalline solids is quantitatively understood through defect-mediated atomic hops, no comparable quantitative framework exists for glasses. In these systems, the origin of large diffusion activation energies remains puzzling, despite local rearrangements involving low barriers. Using molecular dynamics simulations of metallic glasses, we decompose diffusion into random-walk and correlation contributions and find that back-and-forth correlated motion – not local rearrangement barriers – dominates the activation energy, resolving how low-barrier rearrangements yield large macroscopic activation energies. These correlations arise from asymmetry between forward and reverse barriers, a generic feature of disordered energy landscapes. We find that the correlation-driven mechanism is active beyond metallic glass alloys, including $SiO_2$ and a single-component Lennard-Jones glass. The latter demonstrates that the correlation originates from structural disorder rather than chemical complexity. The framework also explains accelerated surface diffusion, where reduced activation energies arise primarily from weaker correlations rather than changes in local rearrangement barriers. Our results establish a direct, quantitative link between atomic-scale dynamics and macroscopic transport, providing a predictive basis for kinetics in disordered materials.

Diffusion is a fundamental and technologically critical property of solids[1], governing microstructural evolution in crystals and aging and transport in glasses, with broad implications for structural materials, semiconductors, batteries, nuclear waste storage, and drug delivery[2–6]. At the atomic-scale, diffusion in solids proceeds via thermally activated rearrangements overcoming an activation barrier. For simple crystals in which diffusion proceeds by a single-hop process, like vacancy-exchange, transition-state theory gives the relationship between these barriers and the macroscopic diffusion coefficient $D$ as $D = \nu a^2 g f x_D \exp\left(\frac{-\Delta E}{k_B T}\right)$, where $\nu$ is the attempt frequency, $a$ is the hop distance, $g$ is the geometric factor, and $x_D$ is the concentration of diffusion-mediating defects[7]. $\Delta E$ is the migration barrier between an initial state and the corresponding saddle point. In this case, the diffusion activation energy $E_D$ equals the migration barrier $\Delta E$, directly linking the energetics of individual atomic hops to macroscopic transport. The correlation factor $f$ accounts for directional correlations between successive atomic hops that reduce net diffusion relative to an ideal random walk and can be computed from lattice geometry[8]. This framework directly links atomic-scale energetics to macroscopic diffusion and underpins predictive materials design in crystalline systems[2–4].

An equally simple atomic-scale description does not exist for glasses. In glasses, structural disorder gives rise to heterogeneous, collective rearrangements spanning wide ranges of displacement lengths and activation barriers[9–11], making it difficult to connect elementary atomic processes to macroscopic diffusion. Experiments and simulations[12–17] show that diffusion in the glassy state, commonly attributed to local atomic rearrangements called β-relaxations[18,19], follows Arrhenius temperature dependence with a well-defined activation energy $E_D$. These β-rearrangements involve the collective displacement of multiple atoms and, through their accumulation over time, give rise to long-range diffusion. However, it remains unclear why $E_D$ is large given that individual β-rearrangement barriers are typically small[20]. In particular, it is unknown whether $E_D$ is determined solely by local rearrangement barriers or whether additional features of the disordered structure such as defect-like excitations or correlated, reversible motion also play a role.

Here we address this question using molecular dynamics (MD) simulations of several model glasses. We introduce a quantitative framework that decomposes the diffusion coefficient as $D = D_{\mathrm{RW}} \times f$, where the random-walk contribution $D_{\mathrm{RW}}$ quantifies the intrinsic propensity for local rearrangements and the correlation factor $f$ measures how effectively these rearrangements produce irreversible net displacement. Applying this decomposition to bulk diffusion in multiple glass-forming systems reveals that correlations exhibit Arrhenius behavior with an associated activation energy $E_f$. This correlation barrier constitutes a substantial fraction of the overall diffusion activation energy $E_D$ in MD-cooled glasses and is predicted to dominate under typical experimental glass-forming conditions. Analysis of the underlying energy landscape suggests that this behavior originates from pronounced asymmetry between forward and reverse rearrangement barriers, a generic feature of glassy landscapes[10,21,22]. Extending the same framework to surface diffusion further shows that the reduced activation energy at free surfaces is governed primarily by a reduction in $E_f$, rather than by uniformly lower local rearrangement barriers. Together these results demonstrate that diffusion in glasses is governed not primarily by the rate of local rearrangements, but by the degree to which these motions escape reversibility, providing a mechanistic understanding of factors controlling glass diffusion that can support connecting atomic rearrangements and macroscale transport coefficients across materials and conditions.

## Results

## Decomposing glassy diffusion into random-walk and correlation contributions

To evaluate the decomposition $D = D_{\mathrm{RW}} \times f$, we compute the tracer diffusion coefficient $D$ from MD simulations using Einstein's relation (see Methods). Fig. S1a shows the mean-squared displacement (MSD) of atoms for glassy $Cu_{50}Zr_{50}$ at 700 K (the glass transition temperature, $T_g$) and 640 K, where long-time diffusive behavior clearly emerges. Over the accessible temperature range, the Arrhenius plot of $D$ (Fig. 1a) yields an activation energy $E_D \approx 1.22$ eV. Systems with broad distributions of local rearrangement barriers need not exhibit simple Arrhenius behavior, yet we observe approximately Arrhenius diffusion here. This is likely a consequence of the limited simulation temperature window and should not be taken as evidence of a single well-defined microscopic barrier. Indeed, simulations reveal a broad distribution of local rearrangement barriers in Cu-Zr and other model glasses, with a substantial fraction at low energies[9,10,23]. Experiments on metallic glasses similarly identify frequent low-barrier localized rearrangements (β relaxations and stress-activated events of ~0.3–0.5 eV)[20], whereas radiotracer diffusion below $T_g$ reports much larger activation energies of ~1–3 eV for long-range diffusion[13]. Reconciling these disparate energy scales is nontrivial. We find that decomposing diffusion into random-walk and correlated contributions resolves this apparent paradox and clarifies the microscopic origin of the diffusion barrier in glasses.

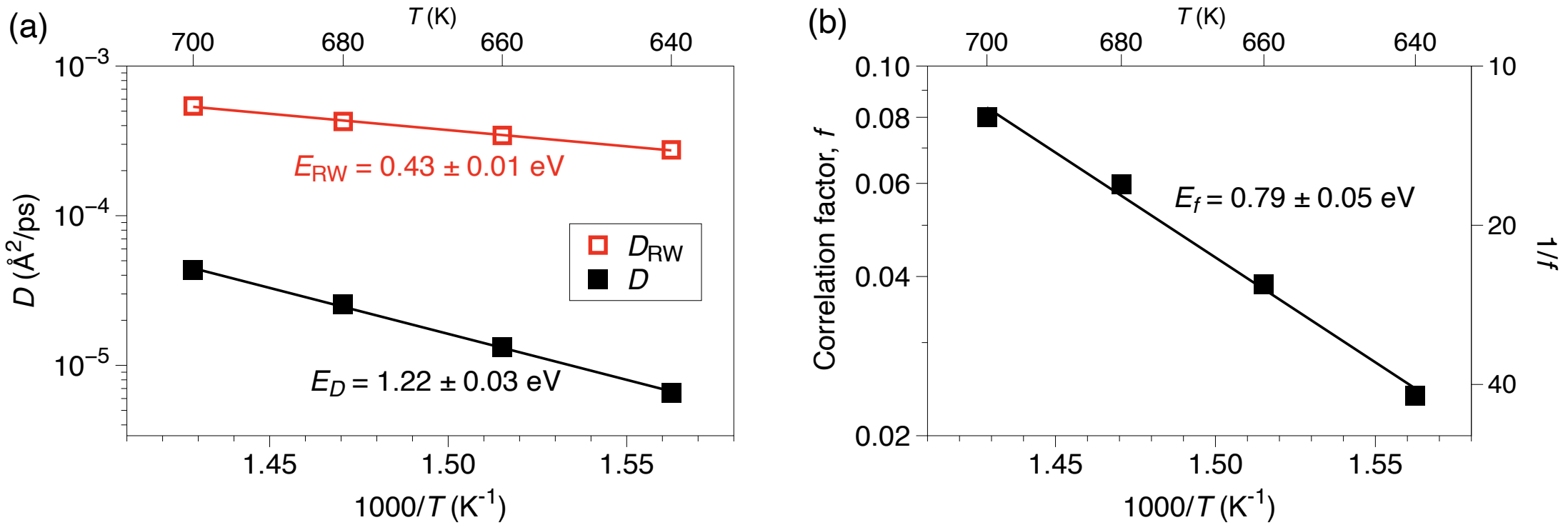

**Fig. 1| Decomposition of glassy diffusion into random-walk and correlation components in $Cu_{50}Zr_{50}$.** Arrhenius plots showing **a,** the total (tracer) diffusion coefficient $D$, and the random-walk diffusion coefficient $D_{\mathrm{RW}}$, and **b,** the correlation factor $f$, at different temperatures in the glassy state. Solid lines are linear fits with the corresponding activation energies indicated. To compute $D_{\mathrm{RW}}$, we identify rearrangements as atomic displacements exceeding 1 Å within a 1 ps interval.

We quantify deviations from ideal random-walk behavior using the correlation factor $f$ defined as:

$$f = D/D_{\mathrm{RW}} \tag{1}$$

where $D_{\mathrm{RW}}$ is the diffusion coefficient of a random-walk trajectory constructed from the atomic displacements in the original MD simulation (see Methods). Smaller values of $f$ indicate stronger temporal correlations (i.e., more pronounced back-and-forth motion). Here, we use "correlation" to refer to memory effects between successive atomic rearrangements, distinct from the spatially cooperative rearrangements also present in our simulations. To compute $D_{\mathrm{RW}}$, we identify activated rearrangements in the MD trajectory and replace each with a step of the same magnitude but with a random direction, while preserving smaller displacements unchanged (Fig. M1 in Methods). This procedure is straightforward in crystalline solids but is more challenging in glasses, where there is no unique criterion for identifying activated rearrangements (Sections S2 and S3). We therefore compare inherent MD configurations sampled every 1 ps and classify atomic displacements exceeding a threshold distance $d$ between successive snapshots as activated rearrangements. This approach gives similar results to the commonly used $D^2_{\text{min}}$ metric[24], as discussed in Methods. Unless otherwise noted, we use $d$ = 1 Å and find that varying $d$ does not affect the qualitative conclusions (Section S4). This random-walk analysis, similar to the tracer diffusion calculations, tracks just one atom at a time. However, the dynamics still includes the full effects of spatially correlated motion in the simulations. Agreement with theoretical predictions in the crystalline limit validates the approach (Section S3).

Figures 1a and 1b show the temperature dependence of $D_{\text{RW}}$ and $f$ (with convergence illustrated in Fig. S1b) for glassy $Cu_{50}Zr_{50}$. Both exhibit clear Arrhenius behavior, with activation energies satisfying $E_D = E_{\text{RW}} + E_f$, consistent with Eq. (1). The correlation factor is significantly smaller than unity, indicating that correlated back-and-forth motion strongly suppresses net diffusion. The associated correlation barrier $E_f$ accounts for ~65% of $E_D$, demonstrating that increasingly correlated motion plays a dominant role in the slowdown of diffusion with temperature in the glassy state. The origin of this large $E_f$ is structural, not chemical, making it fundamentally different from temperature-dependent correlation factors in crystalline solids (see Discussion).

The physical meaning of the small correlation factor is evident at the level of individual atomic trajectories. By ranking atoms according to their long-time displacements at 640 K, we find that both the most and least mobile atoms exhibit comparable random-walk MSDs. However, the ratio of the random-walk MSD to the true MD MSD ranges from ~10 for the most mobile atoms to ~$10^4$ for the least mobile. Thus, atoms that appear immobile over long times undergo frequent local rearrangements, but with strongly correlated motions that largely cancel, yielding little net displacement. In contrast, the most mobile atoms undergo rearrangements that are more often irreversible and therefore contribute efficiently to long-range diffusion. This behavior mirrors experimental observations on glass surfaces far below $T_g$: scanning tunneling microscopy reveals that surface clusters in metallic glasses and amorphous silicon undergo extensive back-and-forth hopping between a small number of metastable states with very little net diffusion[25–27]. This behavior is a direct surface analog of the correlation-dominated dynamics identified in this work, and consistent with the reduced (but still significant) correlation effects we find at free surfaces, as discussed below. Thus, with decreasing temperature, diffusion in glasses is not primarily suppressed by a reduction in the rate of local rearrangements, but rather by the predominance of reversible motion encoded in the correlation factor.

The qualitative results of Fig. 1 are insensitive to the details of the rearrangement-identification procedure (Section S4). Varying the displacement threshold $d$ over a wide range preserves Arrhenius behavior in both $D_{\text{RW}}$ and $f$. Even for $d$ = 2 Å, $E_f$ contributes ~40% of $E_D$, demonstrating that correlated motion remains important even for larger-scale rearrangements. We also identify rearrangements using an alternative approach based on clustering full atomic trajectories to separate dense regions of configuration space (stable states) from sparse regions (transition paths) (Section S4). Random-walk trajectories constructed from this approach again yield Arrhenius behavior with similar relative contributions for both $D_{\text{RW}}$ and $f$, confirming the robustness of the decomposition.

Applying the same analysis to the covalent-stabilized metal-metalloid glass $Ni_{80}P_{20}$ (Section S5) yields similarly low $f$ values, though roughly twice that in $Cu_{50}Zr_{50}$, and the correlation barrier $E_f$ is ~50% of $E_D$ (versus 65% for $Cu_{50}Zr_{50}$). Additional studies on the oxide glass $SiO_2$ and a single-component Lennard-Jones glass show the same trend (Section S5). Together, these results demonstrate that strong, temperature-dependent correlation effects are general across chemically and structurally distinct glass-forming systems and can originate from structural disorder alone without the need for chemical complexity, a key distinction from superficially similar effects in crystalline alloys[28].

### Correlation effects in experimentally relevant glasses

A key question is whether the strong correlations observed here are artifacts of the high cooling rates used in MD simulations (~$10^9$ K/s in the data shown) or persist in glasses formed at experimentally relevant rates of $10^1$–$10^5$ K/s. To address this, we evaluated $f$ and $D$ for glasses quenched at rates from $5\times10^8$ to $10^{12}$ K/s (Section S6) and find that slower cooling leads to smaller $f$, indicating increasingly correlated atomic motion. To estimate correlation factors for experimentally relevant glasses, we exploit a well-known empirical linear relationship that often emerges between the logarithms of different thermally activated kinetic parameters in glasses[17,29]. A plot of log($f$) versus log($D$) displays an almost perfectly linear relationship (Fig. 2). Extrapolating this relation to diffusion coefficients typical of experimental metallic glasses ($D \approx 10^{-22}$ m$^2$/s = $10^{-14}$ Å$^2$/ps)[12,30] yields $f \approx 10^{-7}$ as shown in the inset of Fig. 2. While approximate, this estimate reveals that correlation effects increase dramatically with slower cooling and are likely orders of magnitude stronger in experimentally relevant glasses.

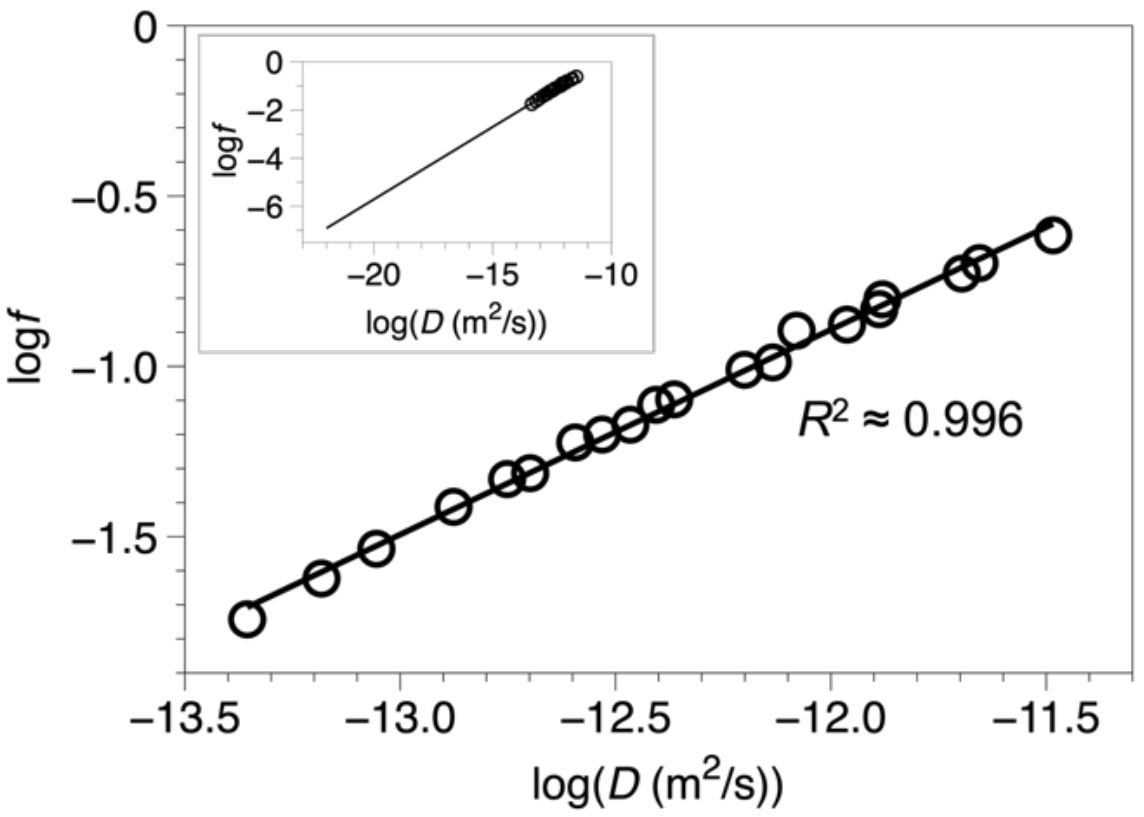

**Fig. 2| Scaling relationship between correlation effects and diffusion coefficient across cooling rates.** Log-log plot of the correlation factor $f$ versus the diffusion coefficient $D$ from simulations at different cooling rates. The linear relationship enables extrapolation to experimentally relevant conditions: the inset shows that for $D = 10^{-22}$ m$^2$/s (typical of experimental metallic glasses[30]), $f \approx 10^{-7}$, indicating that correlation effects dominate by orders-of-magnitude in laboratory-prepared glasses.

### Local barrier distributions explain the random-walk component

Having quantified the decomposition of the diffusion activation energy $E_D$, we now relate it to the underlying potential energy landscape (PEL) by analyzing the contributions to $E_{\mathrm{RW}}$ and $E_f$. If random-walk diffusion is governed by local rearrangement barriers, $E_{\mathrm{RW}}$ should reflect the activation barriers sampled during the dynamics, while $E_D$ should be substantially larger due to correlation effects. To test this hypothesis, we used nudged elastic band (NEB)[31,32] calculations to compute rearrangement barriers $E_{\mathrm{act}}$ for 5,000 randomly sampled rearrangement events occurring within 1 ps between inherent structures explored during MD at 640 K. Figure 3 shows the resulting rearrangement barrier distribution (solid curve) for events with maximum atomic displacement ≥ 1 Å, consistent with the threshold $d$ used to identify activated rearrangements. The distributions shift systematically to higher energies with increasing displacement, consistent with previous observations[33]. For the 1 Å threshold, most barriers lie below 0.5 eV and nearly all below 1 eV, consistent with $E_{RW}$ = 0.43 eV (Fig 1a) but much smaller than the total diffusion barrier $E_D$ = 1.22 eV. Iso-configurational ensemble MD simulations[34] and the Activation-Relaxation Technique yield similar conclusions (Section S7), confirming that these results are independent of the specific sampling method. Also, $E_{RW}$ increases systematically with the displacement threshold (Section S4), tracking the upward shift of the barrier distributions in Fig. 3. Interestingly, $E_{RW}$ is nearly insensitive to cooling rate (Section S6), despite the systematic upward shift of forward-barrier distributions in more slowly quenched glasses. This suggests that random-walk diffusion is controlled by the low-energy tail of the barrier distribution that is thermally accessible, while the high-energy barriers introduced by slower quenching manifest instead in enhanced correlation effects.

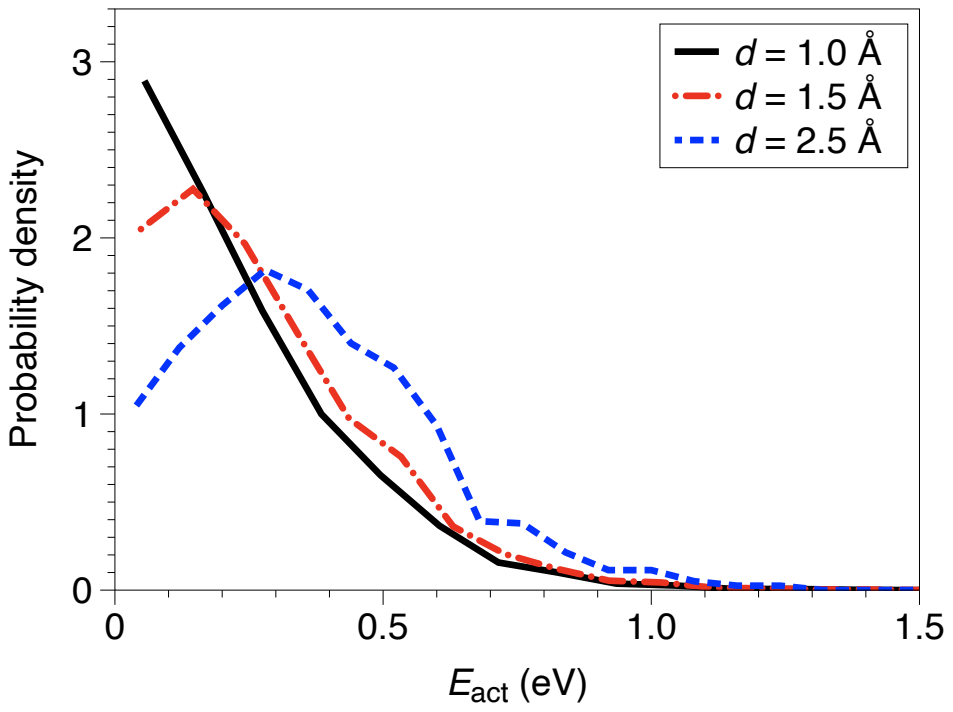

**Fig. 3| Distribution of local rearrangement barriers from nudged elastic band calculations.** Activation barriers for thermally sampled rearrangement events in $Cu_{50}Zr_{50}$ at 640 K, shown for three displacement thresholds (≥ 1 Å, ≥ 1.5 Å, and ≥ 2.5 Å). The distributions shift systemically to higher energies with increasing displacement magnitude. For the 1 Å used throughout this work, the vast majority of barriers lie below 0.5 eV, consistent with $E_{RW}$ = 0.43 eV, and nearly all below 1 eV, much smaller than the total diffusion barrier $E_D$ = 1.22 eV.

### Correlated motion and its relation to the potential energy landscape (PEL)

While local rearrangement barriers govern $E_{RW}$, the correlation barrier $E_f$ reflects the topology of the rugged glassy PEL. In disordered systems, neighboring minima are connected by intrinsically asymmetric barriers, such that the activation energy for a forward rearrangement differs from that of its reverse (Fig. 4a). This asymmetry is a generic feature of disordered energy landscapes[9,10] and provides a microscopic origin for correlated back-and-forth motion: atoms that cross a forward barrier often encounter a much lower reverse barrier and preferentially retrace their steps rather than contribute to net displacement.

We illustrate this asymmetry in $Cu_{50}Zr_{50}$ using the activation–relaxation technique (ART)[35–38] to sample forward ($E_{fw}$) and reverse ($E_{rev}$) activation barriers for rearrangements at 700 K, restricting attention to events with maximum atomic displacement ≥ 1 Å. The forward barriers exhibit a broad distribution, peaked near 1.5 eV (Fig. 4b), consistent with previous studies[9,10,39]. In contrast, the corresponding reverse barrier distribution (Fig. 4c) is strongly skewed toward lower energies, reflecting the same forward-reverse asymmetry reported in amorphous silicon[21], binary Lennard-Jones systems[22], and metallic glasses[9,10]. This disparity provides a direct PEL-based explanation for the correlated motion observed in our simulations. Because transition rates depend exponentially on barrier height, the imbalance between forward and reverse rates becomes increasingly pronounced upon cooling: higher-barrier forward transitions are suppressed more rapidly than lower-barrier reverse transitions. This selective suppression enhances the probability of return relative to escape, explaining the strong decrease of $f$ with decreasing temperature and its approximately Arrhenius behavior.

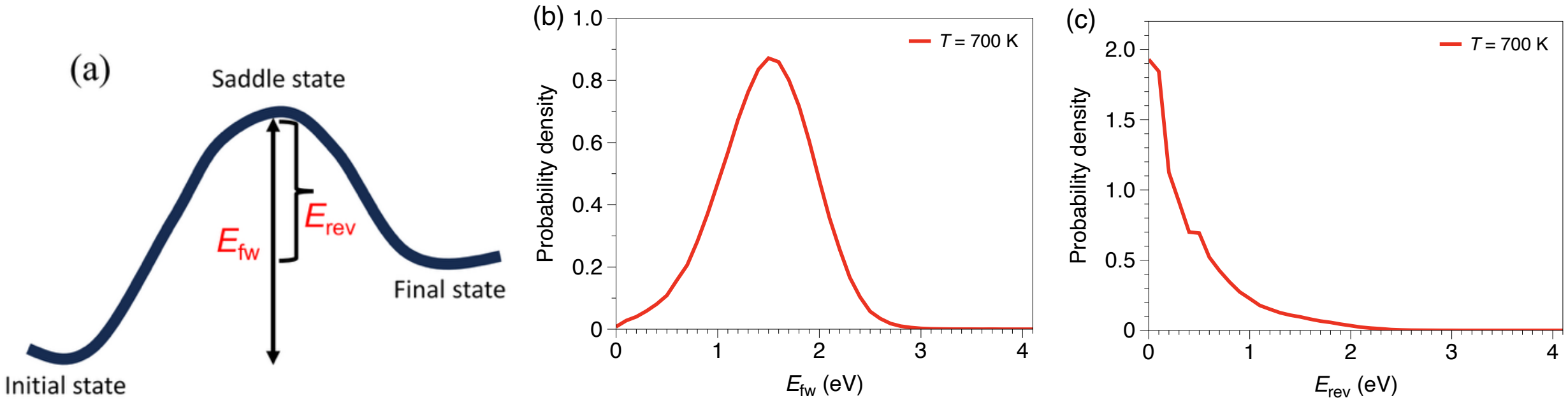

**Fig. 4| Asymmetric potential energy landscape in glasses. a,** Schematic of an asymmetric barrier separating two neighboring minima. The forward transition from initial to final state involves overcoming a forward barrier $E_{\text{fw}}$, while the reverse transition has a distinct barrier $E_{\text{rev}}$, with $E_{\text{fw}} \neq E_{\text{rev}}$. Probability densities of **b,** forward-barriers $E_{\text{fw}}$ and **c,** reverse barriers $E_{\text{rev}}$ in $Cu_{50}Zr_{50}$ at 700 K, obtained using ART. Only events with maximum atomic displacement ≥ 1 Å are included. The reverse barrier distribution is strongly skewed toward lower energies, reflecting the systematic asymmetry that underlies correlated motion in glasses.

The same asymmetry explains the strong cooling-rate dependence of $f$ in our MD simulations (Fig. 2 and Section S6). Previous studies of Cu-Zr metallic glasses[9,10] show that slower quenching shifts forward activation barriers to higher energies while leaving reverse barriers largely unchanged, thereby enhancing barrier asymmetry in more relaxed glasses. Consistent with this picture, Fig. S7 demonstrates that $f$ at a given temperature decreases continuously as the cooling rate is reduced over nearly four orders of magnitude in our simulations. Increased asymmetry at lower cooling rates also strengthens the temperature dependence of $f$ (i.e., increases $E_f$). As forward barriers grow, escape pathways are suppressed more rapidly upon cooling, resulting in a stronger Arrhenius dependence of the correlation factor $f$. MD results (Fig. 5 and Table S1) confirm this behavior: as cooling rate decreases from $10^{12}$ K/s to $5 \times 10^8$ K/s, $E_f$ increases substantially while $E_{\text{RW}}$ remains nearly constant, so that changes in $E_D$ are dominated by $E_f$.

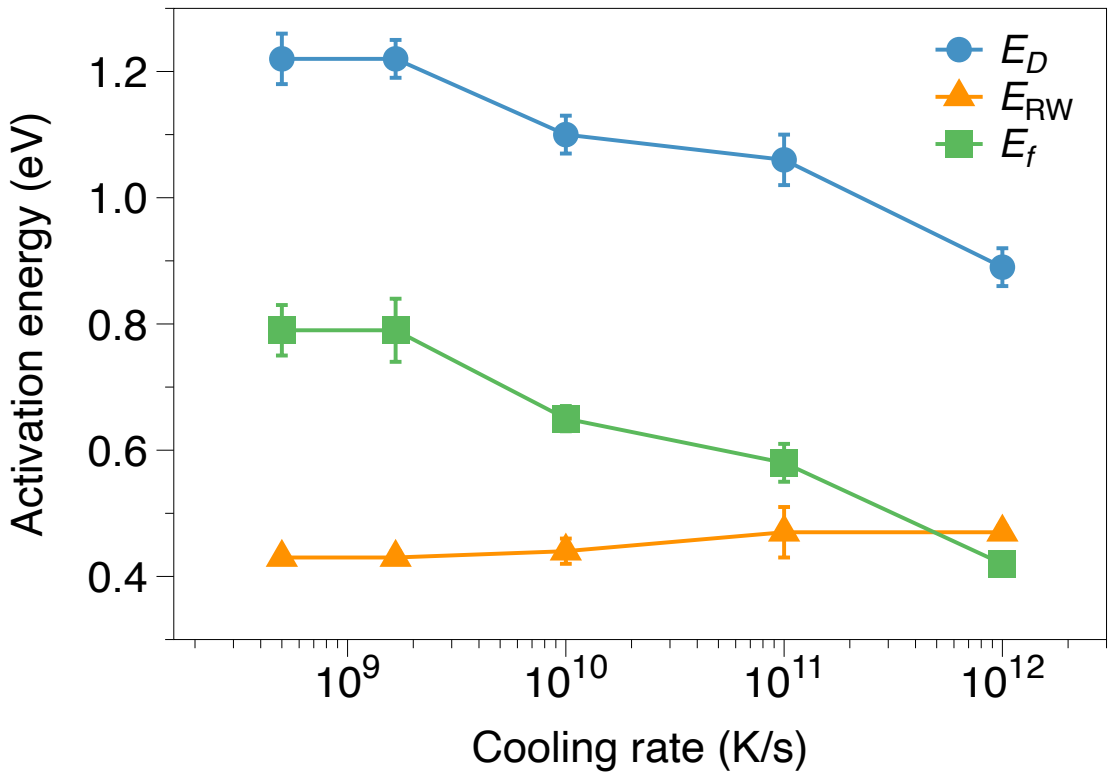

**Fig. 5| Cooling-rate dependence of diffusion activation energies in $Cu_{50}Zr_{50}$.** Activation energies associated with the total diffusion coefficient $E_D$, the random-walk contribution $E_{\text{RW}}$, and the correlation barrier $E_f$ in the glassy state, plotted as a function of cooling rate. While $E_{\text{RW}}$ remains nearly constant across almost four orders of

magnitude in cooling rate, $E_f$ increases substantially with slower quenching, demonstrating that the strong cooling-rate dependence of $E_D$ originates primarily from enhanced correlation effects.

Scanning tunneling microscopy measurements, albeit on surfaces, provide a way to quantify barrier asymmetry in real glasses. By imaging the two-state hopping dynamics of atomic clusters on metallic glass and amorphous silicon surfaces well below $T_g$, Ashtekar et al.[25–27] extracted the free-energy difference between pairs of metastable states – a direct measure of landscape asymmetry at the single-event level. They revealed a broad distribution of asymmetries across different sites, ranging from near-zero (symmetric double wells) to several $k_BT$ (strongly asymmetric), consistent with the spatially heterogeneous barrier asymmetry expected from structural disorder.

**Surface diffusion: reduced correlations rather than reduced barriers**

As a final application of the decomposition in Eq. (1), we analyze surface diffusion in $Cu_{50}Zr_{50}$ and contrast it with bulk behavior. Surface diffusion plays a central role in phenomena such as ultrastable glass formation[40] and is known to be faster and to exhibit lower activation energies than bulk diffusion[11,17,41,42]. By analogy with crystalline metals[43], this enhancement is commonly attributed to reduced local rearrangement barriers at free surfaces[44]. Here, we reassess this picture by explicitly accounting for correlation effects.

Figure 6a shows Arrhenius plots of the surface diffusion coefficient $D_s$ and the corresponding surface random-walk diffusion coefficient $D_{s,\mathrm{RW}}$, extracted from the two-dimensional MSD of surface atoms (Methods). Although the surface MSD is weakly non-Fickian over the accessible timescales (Fig. S11a), consistent analysis across temperatures yields reliable activation energies and well-converged surface correlation factors $f_s = D_s/D_{s,\mathrm{RW}}$ (Fig. S11b). The surface random-walk activation energy (0.35 eV) is slightly lower than in the bulk (0.43 eV), consistent with reduced local rearrangement barriers at free surfaces[11]. However, surface atoms exhibit substantially larger correlation factors (i.e., less correlated motion) at all temperatures (Fig. 6b), indicating markedly less back-and-forth motion. Thus, while random-walk barriers differ only modestly between bulk and surface, correlation barriers differ substantially (0.79 eV versus 0.44 eV). The reduced surface diffusion barrier therefore arises primarily from suppressed correlation effects rather than from uniformly reduced local rearrangement barriers.

The ratio $D_s/D$ near $T_g$ is a key indicator of the potential for ultrastable glass formation. From Eq. (1), we can write:

$$\frac{D_s}{D} = \frac{D_{s,RW}}{D_{RW}}\frac{f_s}{f}. \tag{2}$$

Over the temperature range studied, the dominant contribution to $D_s/D$ comes from $D_{s,\mathrm{RW}}/D_{\mathrm{RW}} \approx 10$. However, this ratio is nearly temperature independent (Figs. S11c and 6c); most of the temperature dependence of $D_s/D$ therefore arises from $f_s/f$, which increases from ~3 at 700 K to ~5.5 at 640 K. We expect the correlation effect to become even larger at lower temperatures, as discussed for bulk diffusion. This mechanism is fundamentally distinct from diffusion enhancement at crystalline surfaces and grain boundaries, where reduced coordination lowers intrinsic migration barriers by decreasing bond breaking[45,46]. In glasses, instead, surface diffusion activation barriers are reduced primarily because altered local topology diminishes the tendency for reverse rearrangements.

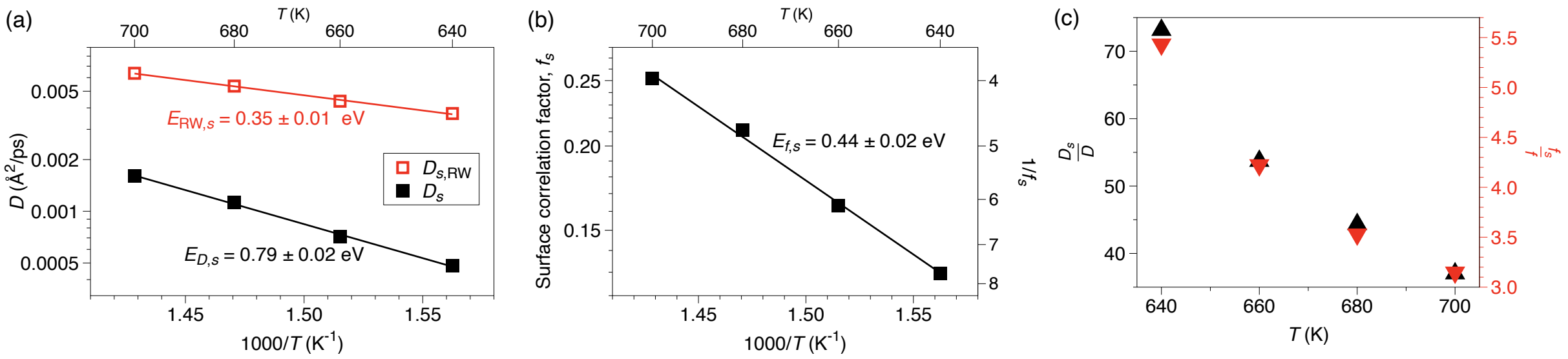

**Fig. 6| Surface diffusion in $Cu_{50}Zr_{50}$ reveals the dominant role of reduced correlations. a,** Temperature dependence of the surface diffusion coefficient $D_s$ and the surface random-walk coefficient $D_{s,\mathrm{RW}}$ (from reconstructed trajectories). **b,** Arrhenius plot of the surface correlation factor $f_s = D_s/D_{s,\mathrm{RW}}$. Solid lines are least-squares fits used to extract activation energies. **c,** Temperature dependence of the surface to bulk diffusion ratio $D_s/D$ and the surface-to-bulk correlation factor ratio $f_s/f$. The random-walk analysis uses a displacement threshold distance of 1 Å over a 1 ps interval.

## Discussion

The asymmetric energy landscape of glasses – where neighboring minima are connected by unequal forward and reverse barriers – provides a fundamental mechanism for correlated atomic motion. This asymmetry produces a systematic bias toward lower reverse barriers, favoring return over escape and strengthening correlations as temperature decreases as observed in different kinds of glasses. Because this mechanism is structural, it persists even in the absence of chemical heterogeneity, as confirmed by similar behavior in a single-component Lennard–Jones glass (Section S5).

In crystalline materials, temperature-dependent correlation factors are only known to emerge from chemical effects. They are well known for dilute solute diffusion[47,48], where they originate from solute-vacancy binding and disparities between solute and host migration barriers[48]. Recent simulations of chemically complex Nb-Mo-Ta alloys[28] show strong temperature-dependent correlations that increase when local migration barriers are widely distributed and diminish in chemically homogeneous systems, recovering the standard lattice value ($f = 0.73$) for pure Nb. These examples show how very significant and temperature-dependent correlation emerges from chemical disorder on a fixed lattice. In contrast, our results demonstrate that in glasses structural disorder alone is sufficient to generate strong correlation effects, though chemical complexity can potentially further amplify them.

The diffusion decomposition established here implies that diffusion-controlled processes in glasses, including aging, stress relaxation, and crystallization, depend not only on the propensity of local rearrangements but also on the likelihood that these rearrangements produce irreversible structural change. The strong cooling-rate dependence of the correlation barrier suggests that the kinetic stability of well-relaxed glasses arises primarily from enhanced reversibility of atomic motion. More generally, large activation energies for kinetic processes in glasses need not reflect uniformly high local barriers but can emerge from asymmetric energy landscapes that favor return over escape. This perspective suggests a potential materials-design strategy: reducing forward–reverse barrier asymmetry could promote irreversible motion and enhance diffusion in amorphous solids. Such control is particularly relevant for fast-ion-conducting glasses, where maximizing diffusion is essential. Conversely, high-entropy metallic glasses, with their wide distribution of atomic environments, may possess large correlation barriers and suppressed diffusion, potentially enhancing radiation tolerance by promoting defect recombination and structural self-healing at elevated temperatures.

## Conclusions

Diffusion in glasses is widely viewed as a collective process reflecting the cooperative dynamics of disordered energy landscapes[49–52]. Our results refine this picture by demonstrating that the dominant contribution to diffusion slowdown at low temperatures arises not primarily from increased barriers to collective rearrangements, but from their growing reversibility, quantified by the correlation factor. By decomposing diffusion into a random-walk contribution associated with local atomic rearrangements and a correlation component reflecting the tendency of these rearrangements to reverse, we show across multiple model glass-formers that a substantial fraction of the total diffusion activation energy originates from dynamical correlations. Extrapolation toward experimentally relevant cooling rates suggests that these correlation effects become even more pronounced in well-relaxed glasses. Within this framework, the random-walk barrier tracks the distribution of local rearrangement barriers, while intrinsic asymmetry between forward and reverse barriers in the glassy potential energy landscape provides a natural microscopic origin for strong, temperature-dependent correlations. These effects are driven by structure, different from related effects in crystals that emerge from chemical disorder. Applying the same analysis to surface diffusion further reveals that reduced surface diffusion activation energy arises predominantly from suppressed correlation effects rather than reduced rearrangement barriers. More broadly, this perspective reframes diffusion in glasses as a competition between activation and irreversibility, providing a unified lens through which to understand transport, relaxation, and kinetic arrest in amorphous materials.

ACKNOWLEDGEMENTS

This work was primarily supported by NSF through the University of Wisconsin Materials Research Science and Engineering Center (DMR-2309000). This work used the TACC's Stampede3 at University of Texas at Austin through allocation TG-MAT240071, from the Advanced Cyberinfrastructure Coordination Ecosystem: Services & Support (ACCESS) program, which is supported by National Science Foundation (NSF) grants #2138259, #2138286, #2138307, #2137603, and #2138296. The authors are also grateful to the Center for High Throughput Computing (CHTC) at UW–Madison for the computing resources.

References:

1. Mehrer, H. *Diffusion in Solids*. (Springer Berlin Heidelberg, Berlin, Heidelberg, 2007). doi:10.1007/978-3-540-71488-0.
2. Ou, X., Sietsma, J. & Santofimia, M. J. Fundamental study of nonclassical nucleation mechanisms in iron. *Acta Mater.* **226**, 117655 (2022).
3. Bracht, H. Diffusion mechanisms and intrinsic point-defect properties in silicon. *MRS Bull.* **25**, 22–27 (2000).
4. Uxa, D., Hüger, E., Meyer, K., Dörrer, L. & Schmidt, H. Lithium-Ion Diffusion in Near-Stoichiometric Polycrystalline and Monocrystalline LiCoO2. *Chemistry of Materials* **35**, 3307–3315 (2023).
5. McCloy, J. S. & Goel, A. Glass-ceramics for nuclear-waste immobilization. *MRS Bull.* **42**, 233–238 (2017).
6. Yu, L. Surface mobility of molecular glasses and its importance in physical stability. *Adv. Drug Deliv. Rev.* **100**, 3–9 (2016).
7. Van der Ven, A., Ceder, G., Asta, M. & Tepesch, P. D. First-principles theory of ionic diffusion with nondilute carriers. *Phys. Rev. B Condens. Matter Mater. Phys.* **64**, (2001).
8. Allnatt, A. R. & Lidiard, A. B. Atomic Transport in Solids. *Atomic Transport in Solids* https://doi.org/10.1017/CBO9780511563904 (1993) doi:10.1017/CBO9780511563904.
9. Fan, Y., Iwashita, T. & Egami, T. Energy landscape-driven non-equilibrium evolution of inherent structure in disordered material. *Nature Communications 2017 8:1* **8**, 1–7 (2017).
10. Ding, J. *et al.* Universal nature of the saddle states of structural excitations in metallic glasses. *Materials Today Physics* **17**, 100359 (2021).
11. Ma, J. *et al.* Fast surface dynamics enabled cold joining of metallic glasses. *Sci. Adv.* **5**, (2019).
12. Tang, X. P., Geyer, U., Busch, R., Johnson, W. L. & Wu, Y. Diffusion mechanisms in metallic supercooled liquids and glasses. *Nature 1999 402:6758* **402**, 160–162 (1999).
13. Faupel, F. *et al.* Diffusion in metallic glasses and supercooled melts. *Rev. Mod. Phys.* **75**, 237 (2003).
14. Lee, D. & Vlassak, J. J. Diffusion kinetics in binary CuZr and NiZr alloys in the super-cooled liquid and glass states studied by nanocalorimetry. *Scr. Mater.* **165**, 73–77 (2019).
15. Shin, D. *et al.* Oxygen tracer diffusion in amorphous hafnia films for resistive memory. *Mater. Horiz.* **11**, 2372–2381 (2024).
16. Annamareddy, A., Voyles, P. M., Perepezko, J. & Morgan, D. Mechanisms of bulk and surface diffusion in metallic glasses determined from molecular dynamics simulations. *Acta Mater.* **209**, 116794 (2021).
17. Annamareddy, A., Li, Y., Yu, L., Voyles, P. M. & Morgan, D. Factors correlating to enhanced surface diffusion in metallic glasses. *J. Chem. Phys.* **154**, 104502 (2021).
18. Yu, H. B., Samwer, K., Wu, Y. & Wang, W. H. Correlation between β Relaxation and Self-Diffusion of the Smallest Constituting Atoms in Metallic Glasses. *Phys. Rev. Lett.* **109**, 95508 (2012).

19. Yu, H.-B., Wang, W.-H. & Samwer, K. The β relaxation in metallic glasses: an overview. *Materials Today* **16**, 183–191 (2013).

20. Yu, H. Bin, Wang, W. H., Bai, H. Y. & Samwer, K. The β-relaxation in metallic glasses. *Natl. Sci. Rev.* **1**, 429–461 (2014).

21. Kallel, H., Mousseau, N. & Schiettekatte, F. Evolution of the potential-energy surface of amorphous silicon. *Phys. Rev. Lett.* **105**, 045503 (2010).

22. Swayamjyoti, S., Löffler, J. F. & Derlet, P. M. Local structural excitations in model glasses. *Phys. Rev. B Condens. Matter Mater. Phys.* **89**, 224201 (2014).

23. Rodney, D. & Schuh, C. Distribution of thermally activated plastic events in a flowing glass. *Phys. Rev. Lett.* **102**, 235503 (2009).

24. Falk, M. L. & Langer, J. S. Dynamics of viscoplastic deformation in amorphous solids. *Phys. Rev. E* **57**, 7192 (1998).

25. Ashtekar, S., Scott, G., Lyding, J. & Gruebele, M. Direct visualization of two-state dynamics on metallic glass surfaces well below T g. *Journal of Physical Chemistry Letters* **1**, 1941–1945 (2010).

26. Ashtekar, S., Lyding, J. & Gruebele, M. Temperature-dependent two-state dynamics of individual cooperatively rearranging regions on a glass surface. *Phys. Rev. Lett.* **109**, 166103 (2012).

27. Ashtekar, S., Scott, G., Lyding, J. & Gruebele, M. Direct imaging of two-state dynamics on the amorphous silicon surface. *Phys. Rev. Lett.* **106**, 235501 (2011).

28. Xing, B., Zou, W., Rupert, T. J. & Cao, P. Vacancy diffusion barrier spectrum and diffusion correlation in multicomponent alloys. *Acta Mater.* **266**, 119653 (2024).

29. Li, Y. *et al.* Surface diffusion is controlled by bulk fragility across all glass types. *Phys. Rev. Lett.* **128**, 075501 (2022).

30. Cao, C. R., Lu, Y. M., Bai, H. Y. & Wang, W. H. High surface mobility and fast surface enhanced crystallization of metallic glass. *Appl. Phys. Lett.* **107**, 141606 (2015).

31. Henkelman, G. & Jónsson, H. Improved tangent estimate in the nudged elastic band method for finding minimum energy paths and saddle points. *J. Chem. Phys.* **113**, 9978–9985 (2000).

32. Henkelman, G. *et al.* A climbing image nudged elastic band method for finding saddle points and minimum energy paths. *J. Chem. Phys.* **113**, 9901–9904 (2000).

33. Liu, C., Guan, P. & Fan, Y. Correlating defects density in metallic glasses with the distribution of inherent structures in potential energy landscape. *Acta Mater.* **161**, 295–301 (2018).

34. Widmer-Cooper, A. & Harrowell, P. Free volume cannot explain the spatial heterogeneity of Debye–Waller factors in a glass-forming binary alloy. *J. Non. Cryst. Solids* **352**, 5098–5102 (2006).

35. Barkema, G. T. & Mousseau, N. Event-based relaxation of continuous disordered systems. *Phys. Rev. Lett.* **77**, 4358–4361 (1996).

36. Beland, L. K., Brommer, P., El-Mellouhi, F., Joly, J.-F. & Mousseau, N. Kinetic activation-relaxation technique. *Phys. Rev. E* **84**, (2011).

37. MacHado-Charry, E. *et al.* Optimized energy landscape exploration using the ab initio based activation-relaxation technique. *Journal of Chemical Physics* **135**, 34102 (2011).

38. Mousseau, N. *et al.* The Activation-Relaxation Technique: ART Nouveau and Kinetic ART. *J. At. Mol. Phys.* **2012**, 925278 (2012).

39. Ding, J. *et al.* Universal structural parameter to quantitatively predict metallic glass properties. *Nature Communications 2016 7:1* **7**, 1–10 (2016).

40. Swallen, S. F. *et al.* Organic Glasses with Exceptional Thermodynamic and Kinetic Stability. *Science (1979).* **315**, 353–356 (2007).

41. Zhu, L. *et al.* Surface self-diffusion of an organic glass. *Phys. Rev. Lett.* **106**, 256103 (2011).

42. Wang, Z. & Perepezko, J. H. Surface diffusion on a palladium-based metallic glass. *Appl. Phys. Lett.* **124**, 91601 (2024).

43. Wang, X. *et al.* Atomistic processes of surface-diffusion-induced abnormal softening in nanoscale metallic crystals. *Nature Communications 2021 12:1* **12**, 5237- (2021).

44. Stevenson, J. D. & Wolynes, P. G. On the surface of glasses. *Journal of Chemical Physics* **129**, 234514 (2008).

45. Balluffi, R. W., Allen, S. M. & Carter, W. C. *Kinetics of Materials*. *Kinetics of Materials* (John Wiley and Sons, 2005). doi:10.1002/0471749311.

46. Schweizer, P. *et al.* Atomic scale volume and grain boundary diffusion elucidated by in situ STEM. *Nature Communications 2023 14:1* **14**, 7601- (2023).

47. Le Claire, A. D. Solute diffusion in dilute alloys. *Journal of Nuclear Materials* **69–70**, 70–96 (1978).

48. Tucker, J. D., Najafabadi, R., Allen, T. R. & Morgan, D. Ab initio-based diffusion theory and tracer diffusion in Ni–Cr and Ni–Fe alloys. *Journal of Nuclear Materials* **405**, 216–234 (2010).

49. Adam, G. & Gibbs, J. H. On the Temperature Dependence of Cooperative Relaxation Properties in Glass-Forming Liquids. *J. Chem. Phys.* **43**, 139–146 (1965).

50. Heuer, A. Exploring the potential energy landscape of glass-forming systems: from inherent structures via metabasins to macroscopic transport. *Journal of Physics: Condensed Matter* **20**, 373101 (2008).

51. Starr, F. W., Douglas, J. F. & Sastry, S. The relationship of dynamical heterogeneity to the Adam-Gibbs and random first-order transition theories of glass formation. *J. Chem. Phys.* **138**, 12A541 (2013).

52. Zhang, H. *et al.* Role of string-like collective atomic motion on diffusion and structural relaxation in glass forming Cu-Zr alloys. *Journal of Chemical Physics* **142**, 164506 (2015).

53. Mendelev, M. I. *et al.* Development of suitable interatomic potentials for simulation of liquid and amorphous Cu–Zr alloys. *Philosophical Magazine* **89**, 967–987 (2009).

54. van Beest, B. W. H., Kramer, G. J. & van Santen, R. A. Force fields for silicas and aluminophosphates based on ab initio calculations. *Phys. Rev. Lett.* **64**, 1955–1958 (1990).

55. Plimpton, S. Fast Parallel Algorithms for Short-Range Molecular Dynamics. *J. Comput. Phys.* **117**, 1–19 (1995).

56. Rapaport, D. C. *The Art of Molecular Dynamics Simulation*. (Cambridge University Press, 2004). doi:10.1017/CBO9780511816581.

57. Avila, K. E., Küchemann, S., Alhafez, I. A. & Urbassek, H. M. Shear-Transformation Zone Activation during Loading and Unloading in Nanoindentation of Metallic Glasses. *Materials 2019, Vol. 12, Page 1477* **12**, 1477 (2019).

## Methods

**Glass preparation and simulation details.** Our primary model system was the binary metallic glass former $Cu_{50}Zr_{50}$, with atomic interactions described by the embedded atom method (EAM) potential from reference[53]. In the Supplementary Information, we validate the main results using a second binary metallic glass, $Ni_{80}P_{20}$, as well as an oxide glass ($SiO_2$) and a single-component Lennard-Jones (LJ) glass.

For the Cu-Zr and Ni-P systems, each containing 16384 atoms, we first equilibrated the liquid at 2000 K for 2 ns. The melts were then quenched to 1000 K at a rate of ~0.01 K/ps, followed by further cooling to 300 K at a slower rate of ~0.001 K/ps, using molecular dynamics (MD) simulations in the NPT ensemble. From the temperature dependence of the simulation volume, we determined that the Cu-Zr (Ni-P) system fell out of equilibrium at approximately 700 K (580 K), which we identify as the glass transition temperature, $T_g$. In this study, we focus on the glassy state in the temperature range 700 – 640 K for $Cu_{50}Zr_{50}$ and 580 – 520 K for $Ni_{80}P_{20}$. To investigate the dynamics at these temperatures, we took the final configuration obtained after quenching to 300 K and annealed it at the target temperature for 1 ns in the NPT ensemble. Property calculations were then performed in the NVT ensemble. Annealing from 300 K to the temperature of interest produced negligible aging over the simulated timescales compared to configurations obtained directly during the quenching process (Fig. S12). Note that our previous studies[16,17] employed configurations obtained directly during quenching.

For amorphous silica, we melted a 3000-atom system at 4000 K for 0.5 ns, then quenched it to 1000 K in 100 K intervals at a rate of 0.04 K/ps. The widely-used Beest–Kramer–van Santen (BKS) potential was employed[54]. We identified $T_g \approx 2600$ K and focus on the glassy state in the temperature range 2500 – 2200 K. For the single-component LJ system, we simulated 2048 atoms at constant density ($\rho = 0.78$). The system was first equilibrated in the liquid state at $T = 1.5$ for $10^6$ steps and then quenched rapidly to $T = 0.2$ over $5 \times 10^5$ steps to form a glass. It was subsequently equilibrated for an additional $10^6$ steps at this temperature before being annealed to higher temperatures in the range 0.32–0.4 to evaluate diffusion; no crystalline order was observed. A timestep of 1 fs was used in all simulations except for the LJ system, where a timestep of 0.002 (in LJ units) was employed.

To evaluate surface diffusion in $Cu_{50}Zr_{50}$, we started from the annealed bulk configuration obtained after the 1 ns NPT run at the target temperature. Free surfaces were introduced by extending the simulation cell by 10 Å along the ±z direction, creating vacuum regions on both sides of the sample. The system was then equilibrated for an additional 1 ns at constant volume to allow full structural relaxation of the newly created surfaces prior to data collection. Following the approach described in reference[16], atoms located within 2.5 Å of each free surface (corresponding to the outermost 12.5 Å when accounting for the expanded simulation cell) were identified as surface atoms, and all surface diffusion properties were computed using this subset. All simulations were performed using the LAMMPS simulation package[55]. All results presented in this paper are based on analyses of inherent structures, obtained by energy minimization of instantaneous MD configurations using the conjugate-gradient algorithm implemented in LAMMPS, and aggregated over four independent simulations.

**Bulk and surface diffusion coefficients.** The (tracer) bulk diffusion coefficient $D$ was calculated using the standard Einstein's equation[56] given by:

$$D = \lim_{t\to\infty} \frac{1}{6Nt} \langle \sum_{i=1}^{N} |\boldsymbol{r}_i(t) - \boldsymbol{r}_i(0)|^2 \rangle = \lim_{t\to\infty} \frac{MSD(t)}{6t} \tag{4}$$

where $N$ is the number of atoms, $\boldsymbol{r}_i(t)$ is the three-dimensional position vector of atom $i$ at time $t$, and $\langle\cdot\rangle$ denotes an ensemble average or an average over multiple time origins. The surface diffusion coefficient $D_s$ was evaluated analogously but using only the in-plane atomic displacements parallel to the free surface:

$$D_s = \lim_{t\to\infty} \frac{1}{4Nt} \langle \sum_{i=1}^{N'} |\tilde{\boldsymbol{r}}_i(t) - \tilde{\boldsymbol{r}}_i(0)|^2 \rangle, \tag{5}$$

where $\tilde{\boldsymbol{r}}_i(t) = (x_i(\mathrm{t}), y_i(\mathrm{t}))$ is the two-dimensional position vector of atom $i$ projected on the $xy$-plane, and $N'$ is the number of surface atoms.

**Random-walk diffusion ($D_{RW}$) from MD simulations.** We now describe the procedure used to evaluate the random-walk diffusion coefficient ($D_{RW}$) from MD trajectories. In amorphous glasses, the rugged potential energy landscape (PEL) leads to a continuous spectrum of atomic displacement magnitudes, as illustrated in Fig. S2*b* for a $Cu_{50}Zr_{50}$ metallic glass. Larger displacements (typically those exceeding a reasonable fraction of the nearest neighbor spacing) correspond to inter-basin, thermally activated atomic hops. These hops inevitably induce localized relaxations in surrounding atoms, giving rise to short-range displacements that reflect the cooperative nature of the rearrangement. However, short displacements can also arise from purely intra-basin motion, which exhibits significant back-and-forth correlations without contributing appreciably to net diffusion. Because short-range displacements in inherent structure trajectories can originate from either mechanism, it is not immediately clear how they should be classified when identifying activated events for evaluating the random-walk diffusion coefficient. Here, we adopt a simple threshold-based approach: an atom is classified as rearranging if its displacement exceeds a chosen cutoff distance. A common alternative for identifying rearranging atoms in simulations of glasses under plastic flow is the non-affine squared displacement ($D^2_{min}$)[24], which isolates the non-affine component of motion and is often interpreted as signaling genuine rearrangements. In practice, however, using $D^2_{min}$ also requires an arbitrary cutoff[57]. Moreover, in $Cu_{50}Zr_{50}$ under equilibrium conditions without plastic flow, we find that $D^2_{min}$ is linearly correlated with the squared displacement (with $R^2 \approx 0.90$), indicating that the two metrics contain very similar information. We therefore use a simple atomic displacement threshold rather than a less intuitive $D^2_{min}$ cutoff.

The MD simulations are first discretized into regular time intervals, and the displacement of each atom is evaluated over each interval. To construct the random-walk trajectory, atomic displacements exceeding the chosen cutoff distance ($d$) are identified and treated as barrier-crossing (activated) events. For each such displacement, a corresponding random-walk step of the same magnitude but with a randomized direction is generated, as illustrated in Fig. M1. The choice of the time interval is important: overly long intervals may miss rapid back-and-forth displacements or combine multiple independent rearrangements into a single step, leading to an underestimation of $D_{RW}$. In this work, we use a time interval of 1 ps, which we consider to be sufficiently short to capture the majority of relevant displacements. Atomic displacements smaller than the cutoff $d$ are treated as they occur in the original MD trajectories when constructing the random-walk paths (Fig. M1). With this definition, the correlation factor $f$ approaches unity as $d$ is increased to large values. The random-walk diffusion ($D_{RW}$) is then computed from the resulting random-walk trajectories using Einstein's relation:

$$D_{\mathrm{RW}} = \lim_{t\to\infty} \frac{1}{6Nt} \langle \sum_{i=1}^{N} \left|\boldsymbol{r}_{i,RW}(t) - \boldsymbol{r}_{i,RW}(0)\right|^2 \rangle, \tag{6}$$

where $\boldsymbol{r}_{i,RW}(t)$ denotes the position of atom $i$ in the random-walk trajectory at time $t$. For surface atoms, the random-walk trajectory is constructed by randomizing the direction of $\tilde{\boldsymbol{r}}_i$ for the atoms with displacements exceeding $d$, and the random-walk diffusion is evaluated as:

$$D_{S,\mathrm{RW}} = \lim_{t\to\infty} \frac{1}{4N't} \langle \sum_{i=1}^{N'} \left| \tilde{\boldsymbol{r}}_{i,RW}(t) - \tilde{\boldsymbol{r}}_{i,RW}(0) \right|^2 \rangle, \quad (7)$$

where $N'$ is the number of surface atoms. All reported values of random-walk diffusion are obtained by averaging over 50 independent random-walk realizations generated from a given MD trajectory. Once $D_{\mathrm{RW}}$ is determined, the correlation factor $f$ is calculated using Eq. 1.

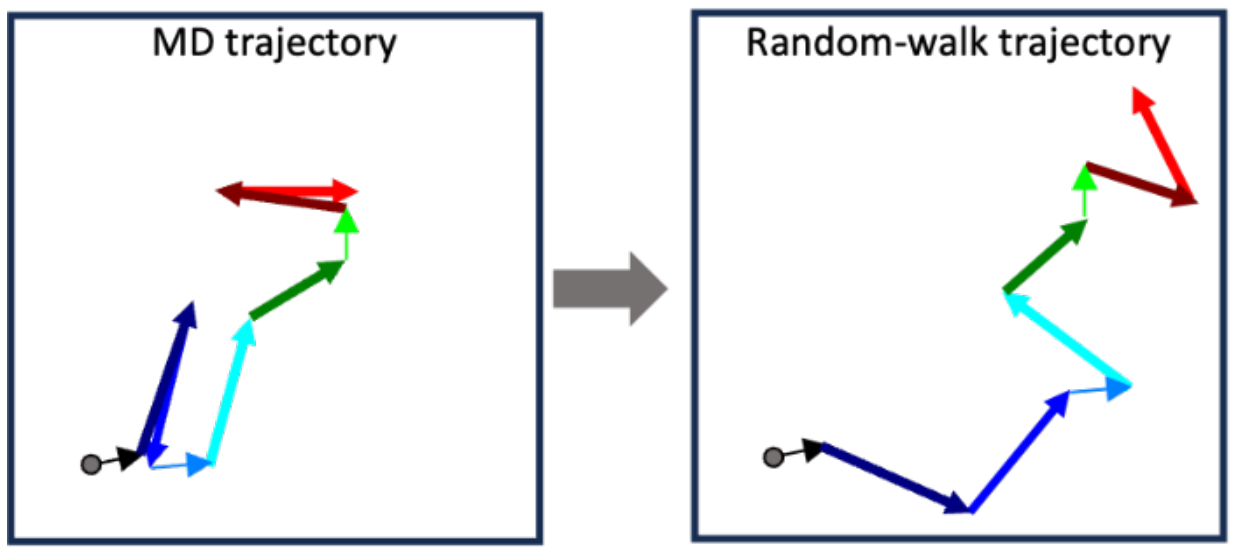

**Fig. M1 | Schematic illustrating the procedure used to compute the random-walk diffusion coefficient ($D_{\mathrm{RW}}$).** The left panel shows a representative two-dimensional atomic trajectory from an MD simulation, where each arrow denotes the displacement vector connecting an atom's initial and final positions over a given time interval. Displacement vectors shorter than the chosen displacement threshold $d$ are shown as thin arrows, whereas displacements exceeding $d$ are shown as thick arrows. The right panel presents the corresponding random-walk trajectory for the same atom, where displacement vectors longer than $d$ are replaced by random-walk steps of identical magnitude but randomized direction, while shorter displacement vectors are retained exactly as observed in the MD simulation. When evaluating $D_{\mathrm{RW}}$ in crystalline systems, the random walk is constrained to the symmetry-equivalent lattice sites accessible to the atom. For example, in FCC nickel, each random-walk step is directed randomly toward one of the 12 nearest-neighbor lattice sites.

**Activation-Relaxation Technique (ART).** We now discuss how we identified rearrangement events and their corresponding barriers in the MD glass configurations. For this purpose, we used the open-source activation-relaxation technique (ART) code[35–38], an open-ended saddle point search algorithm. In ART, an initial perturbation is introduced by randomly displacing a central atom and all atoms within a specified radius. The algorithm then drives the configuration towards a neighboring saddle point in the 3$N$-dimensional configuration space, determining both the activation energy and the final minimum-energy configuration. In this study, we performed 100 ART calculations for each atom in our MD configuration as the *central* atom, using a random initial perturbation for each. Consistent with previous observations in $Cu_{56}Zr_{44}$[33], some identified events exhibited very low activation energies with negligible atomic displacements. After excluding failed and duplicate events, we obtained the forward and revere barrier distributions shown in Figs. 4b and 4c.

## S1. Time evolution of mean-squared displacement (MSD) and correlation factor (*f*) in $Cu_{50}Zr_{50}$:

Fig. S1a shows the time variation of atomic mean-squared displacement (MSD) at two temperatures in the glassy state. At long times, the diffusive behavior is apparent. In Fig. S1b, we show the time-evolution of the inverse of the correlation factor illustrating the convergence of the correlation factor $f$ during our simulations.

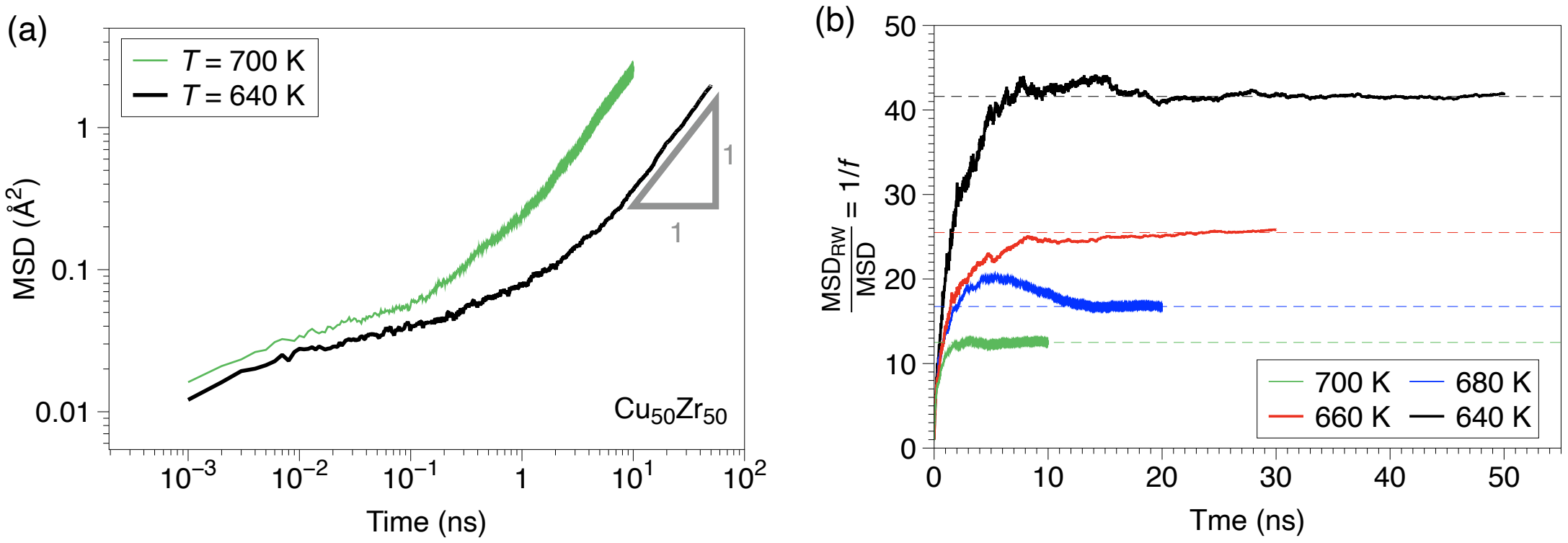

**Fig. S1** | (a) Variation of the mean-squared displacement (MSD) of atoms versus time in $Cu_{50}Zr_{50}$ at two different temperatures within the glassy state. Diffusive motion is observed at longer time scales, as indicated by the log(MSD) versus log(time) plot approaching a slope of unity. (b) Convergence of the correlation factor $f$. For clarity, we plot the time evolution of the ratio of the random-walk MSD to the MD MSD, which by definition equals $1/f$. The ratio approaches a constant value at long times, demonstrating convergence of the calculated correlation factor.

S2. Range of atomic displacements in crystals and metallic glasses:

In crystalline materials, inherent configurations, free of atomic vibrations, reveal a clear separation of length scales between small-amplitude positional fluctuations and activated atomic hops between lattice sites, as illustrated in Fig. S2a, and only these barrier-crossing hops contribute to long-range diffusion. In contrast, in amorphous glasses, the rugged potential energy landscape (PEL) leads to a continuous spectrum of atomic displacement lengths, as illustrated in Fig. S2b for a $Cu_{50}Zr_{50}$ metallic glass obtained from MD simulations.

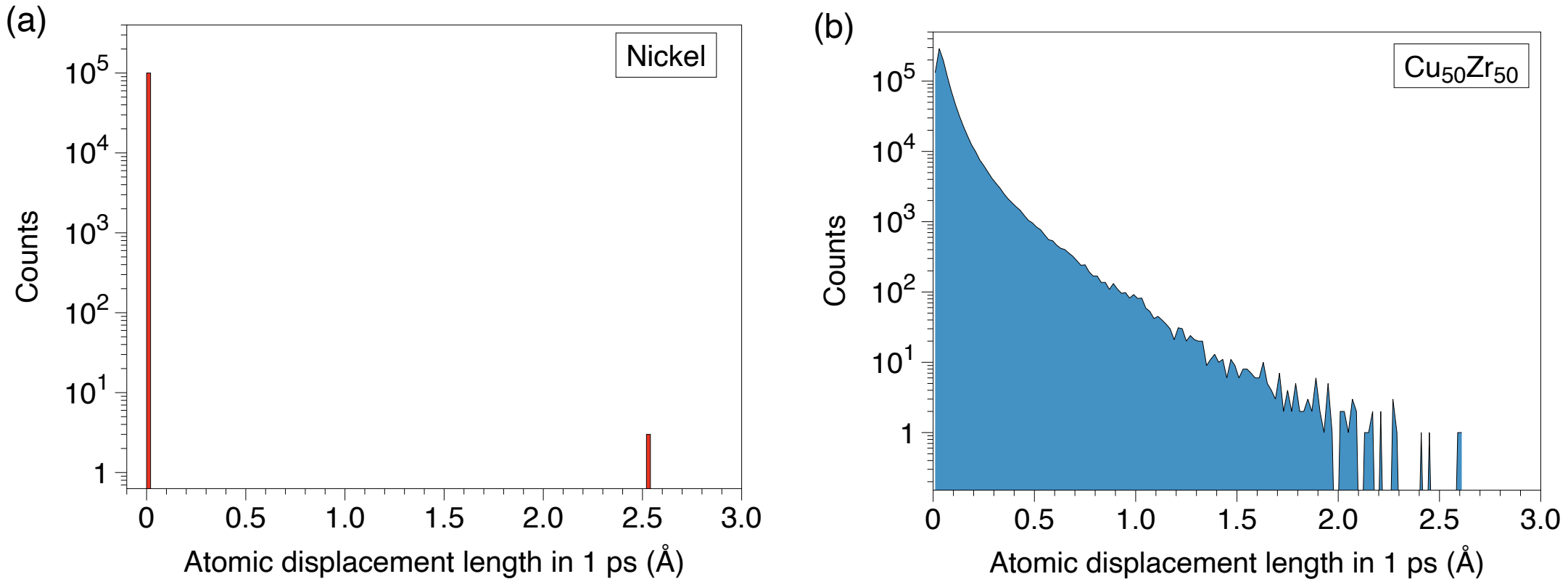

**Fig. S2 |** Histograms showing the distribution of atomic displacement lengths over a 1 ps time interval in (a) crystalline nickel with a single vacancy at 1400K, and (b) amorphous $Cu_{50}Zr_{50}$ at 640 K. The displacements are calculated using inherent configurations. The data reveals a continuous range of displacement lengths in amorphous solids.

## S3. Calculation of correlation factor ($f$) for vacancy-exchange mechanism in nickel:

We have performed MD simulations of FCC nickel containing 1371 atoms and a single vacancy, using an EAM interatomic potential (1). Fig. S3a illustrates the diffusive nature of atomic motions in nickel, with log(MSD) versus log(time) curves approaching a slope of unity at long timescales. Fig. S3b shows the temperature dependence of MD diffusion coefficient $D$ and the corresponding random-walk diffusion coefficient $D_{RW}$. Inherent configurations extracted from MD trajectories at 1 ps intervals are used to evaluate atomic displacements (and shown in Fig. S2a) and identify vacancy-mediated, activated atomic hops for the calculation of $D_{RW}$. The activation energies obtained from Arrhenius fits to $D$ and $D_{RW}$ are nearly identical, resulting in a temperature-independent correlation factor $f \approx 0.76$, in good agreement with the theoretical prediction of 0.78 (2). The simulations were performed over a narrow temperature range chosen to balance two requirements: temperatures high enough to yield reliable diffusion statistics within reasonable simulation times, yet low enough that the maximum displacement of the vacancy within any interval corresponds to a single nearest-neighbor hop. This latter condition ensures that individual activated hops are resolved cleanly, without combining multiple independent hops into a single event or missing rapid back-and-forth vacancy jumps.

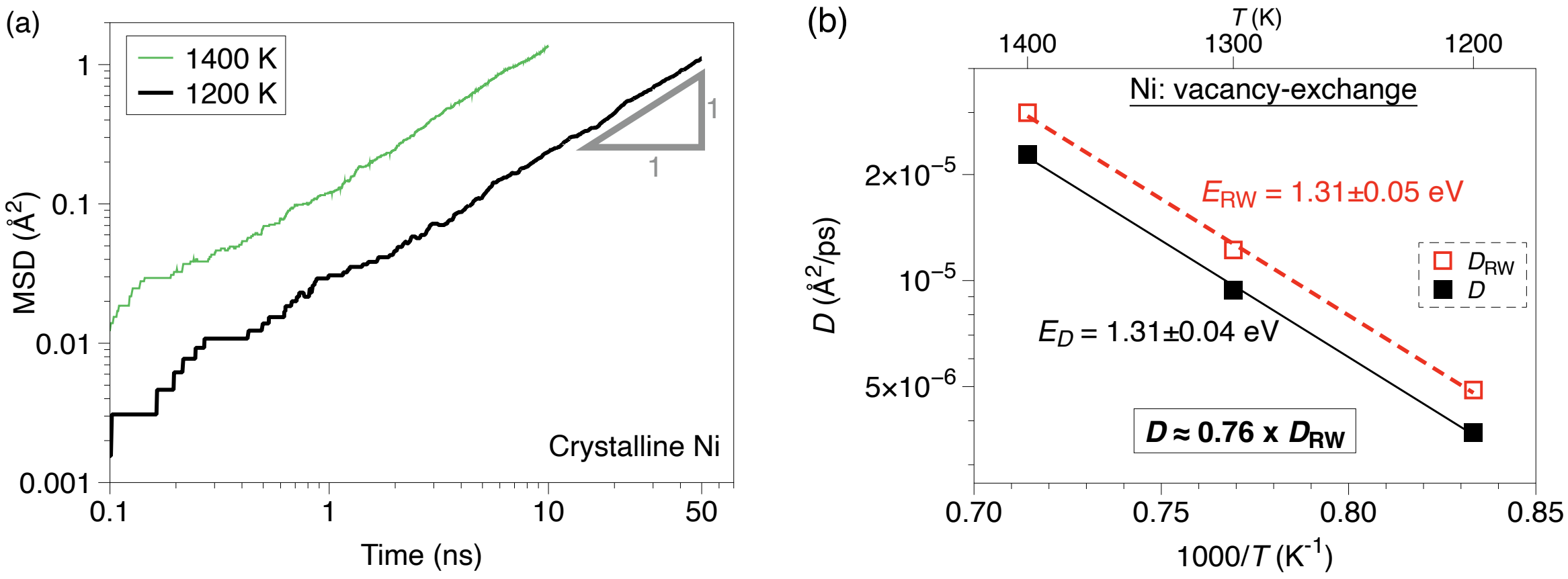

**Fig. S3** | (a) Time evolution of the MSD of atoms in crystalline nickel at the two end temperatures. At long times, diffusive motion is observed, as indicated by the slope of the log(MSD) versus log(time) curves approaching unity. (b) Arrhenius plot of MD diffusion coefficient ($D$) and random-walk diffusion coefficient ($D_{RW}$) against temperature. The lines represent the best linear-fit and the corresponding activation energies are also listed.

S4. Displacement threshold criterion and clustering-based approach for identifying activated hops:

We first begin with the displacement threshold approach. Figs. S4a and S4b show the temperature dependence of the random-walk diffusion coefficient $D_{RW}$ and the correlation factor $f$ for different choices of the displacement threshold $d$ used to identify activated events. The solid lines are Arrhenius fits to the data, and the corresponding activation energies are indicated. In all cases, both $D_{RW}$ and $f$ exhibit clear Arrhenius behavior. The activation energies extracted for different values of $d$ are summarized in Table S1.

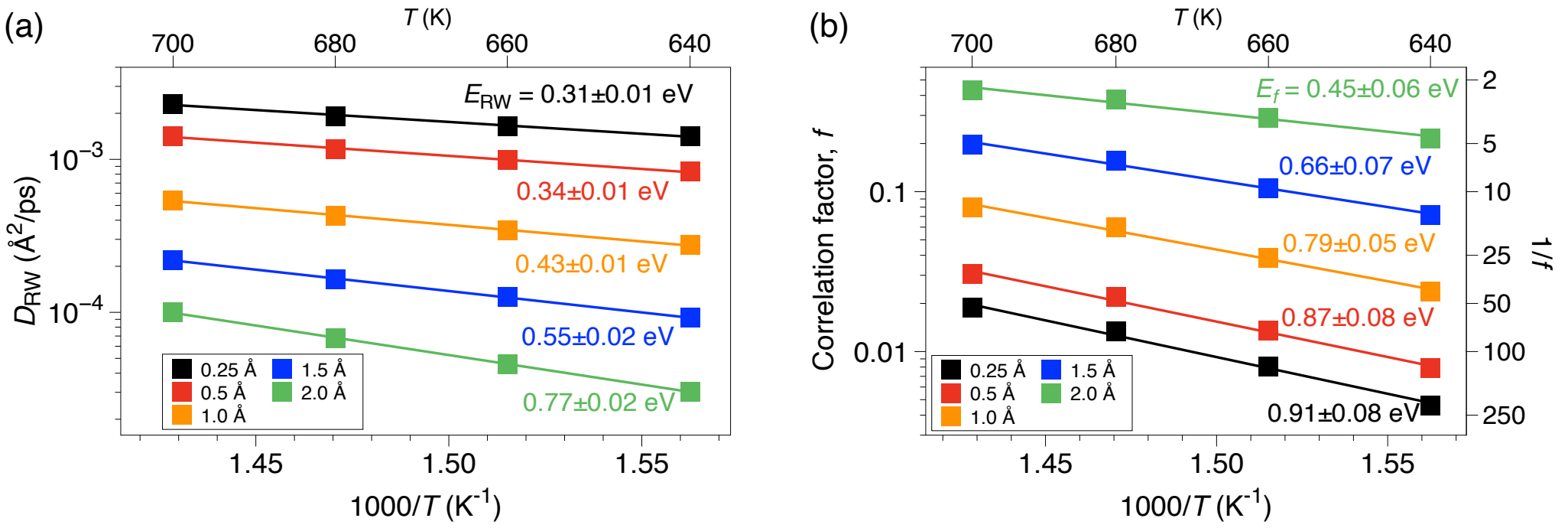

**Fig. S4 |** Arrhenius plots of (a) the random-walk diffusion coefficient $D_{RW}$ and (b) the correlation factor $f$ in the glassy state of $Cu_{50}Zr_{50}$ for different choices of the displacement threshold $d$.

| Displacement threshold, $d$ (Å) | $E_{RW}$ (eV) | $E_f$ (eV) |
|---|---|---|
| 0 | 0.26 | 0.96 |
| 0.1 | 0.28 | 0.94 |
| 0.25 | 0.31 | 0.91 |
| 0.5 | 0.34 | 0.87 |
| 0.75 | 0.39 | 0.83 |
| 1.0 | 0.43 | 0.79 |
| 1.25 | 0.49 | 0.73 |
| 1.5 | 0.55 | 0.66 |
| 1.75 | 0.64 | 0.57 |
| 2.0 | 0.77 | 0.45 |

**Table S1 |** Arrhenius plots of (*a*) the random-walk diffusion coefficient $D_{RW}$ and (b) the correlation factor $f$ in the glassy state of $Cu_{50}Zr_{50}$ for different choices of the displacement threshold $d$.

As an alternative approach for identifying activated hops, we employ the density-based clustering algorithm DBSCAN to group each atom's positions (sampled at 1 ps intervals) into distinct configurational states over the full simulation trajectory. DBSCAN is well suited for the task because it naturally identifies dense regions of configurational space corresponding to stable local states, separated by sparse regions associated with transition pathways. This allows genuine barrier-crossing events to be distinguished from large-amplitude fluctuations within a single basin. Each cluster thus represents a region of the potential energy landscape where the atom resides without crossing a significant energy barrier, and a transition between clusters along the trajectory is identified as an activated hop. As with displacement-based criteria, this method requires the choice of a parameter (the DBSCAN radius, ε) which defines the maximum distance between configurations considered to belong to the same basin. An additional and important issue is the choice of the trajectory length over which clustering is performed to determine whether an atom has undergone a basin change at a given time. In principle, over sufficiently long times an atom's trajectory may span a large region of configuration space (or even the entire simulation box), causing distinct basins to be merged and genuine activated hops to be missed. In the present work, clustering is performed using the full simulation trajectory; however, resolving this timescale dependence is essential if DBSCAN is to be used reliably for identifying basin changes. Using $\varepsilon = 0.5$ Å, Fig. S5a shows the trajectories of seven representative atoms over 5 ns, colored by their cluster assignments, with the number of identified clusters ($n_C$) indicated for each atom. Fig. S5b presents the distribution of atomic jump lengths associated with cluster-to-cluster transitions, showing that most identified hops involve displacements of 1 Å or larger. Notably, approximately 25% of atomic displacements exceeding 1 Å are classified as activated events under this clustering-based analysis. Random-walk trajectories constructed from hops identified using this method (results not shown) also exhibit Arrhenius behavior for both $D_{RW}$ and $f$ over a range of ε values.

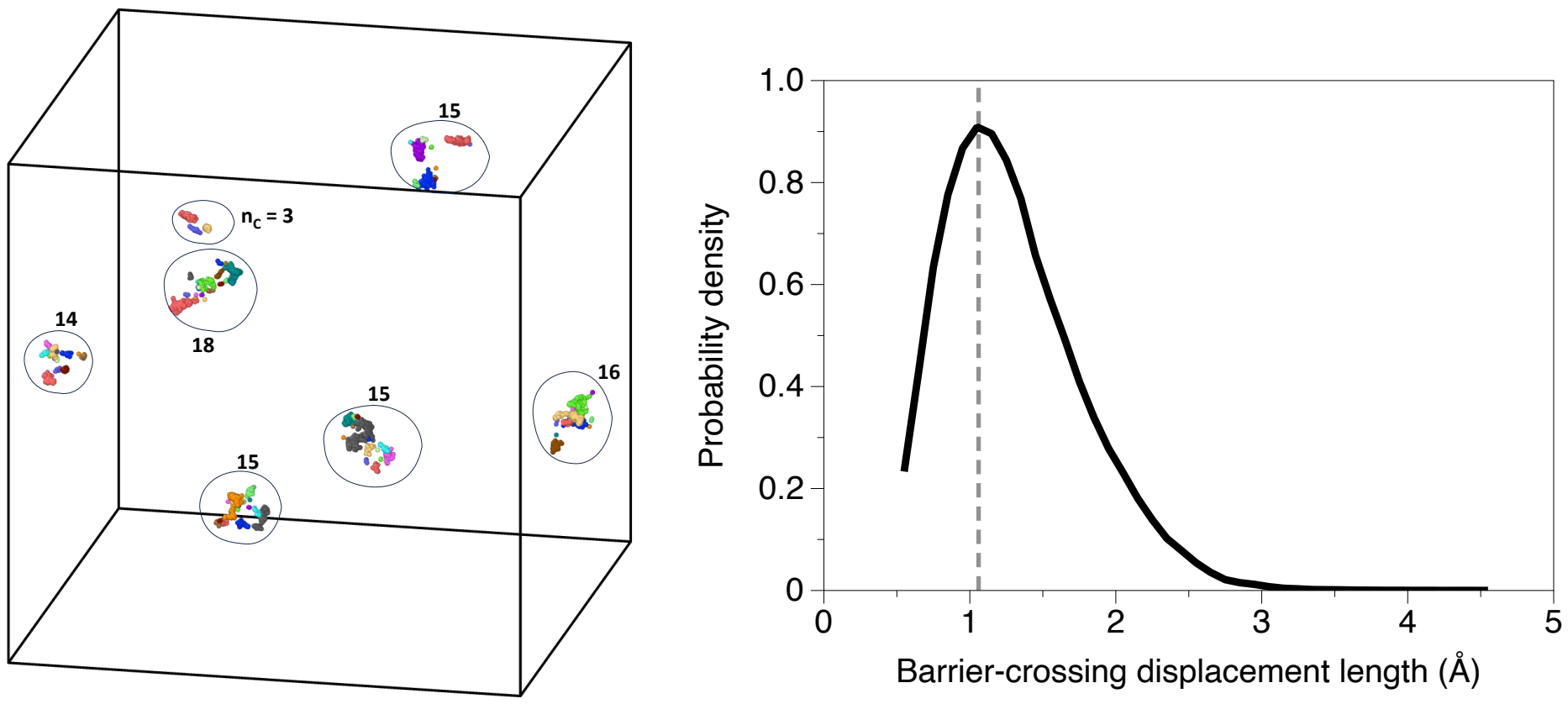

**Fig. S5 |** (a) Trajectories of seven representative atoms over a 5 ns interval, sampled at 1 ps resolution and colored by their DBSCAN cluster assignments. (b) Distribution of atomic displacement lengths associated with barrier-crossing events identified using the DBSCAN method.

## S5. Validation with other glass-formers:

Below we demonstrate the decomposition $D = D_{RW} \times f$ for three different glassy systems: a metallic glass ($Ni_{80}P_{20}$), an oxide glass ($SiO_2$), and a Lennard-Jones (LJ) glass.

$Ni_{80}P_{20}$ glass

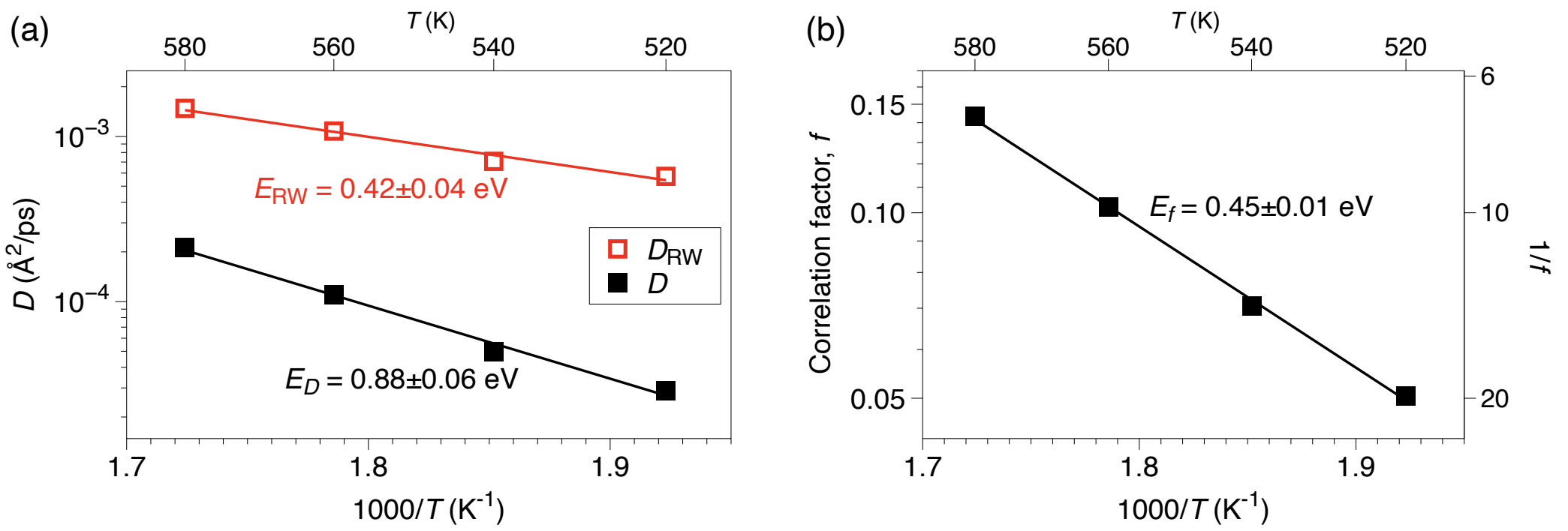

Amorphous silica ($SiO_2$)

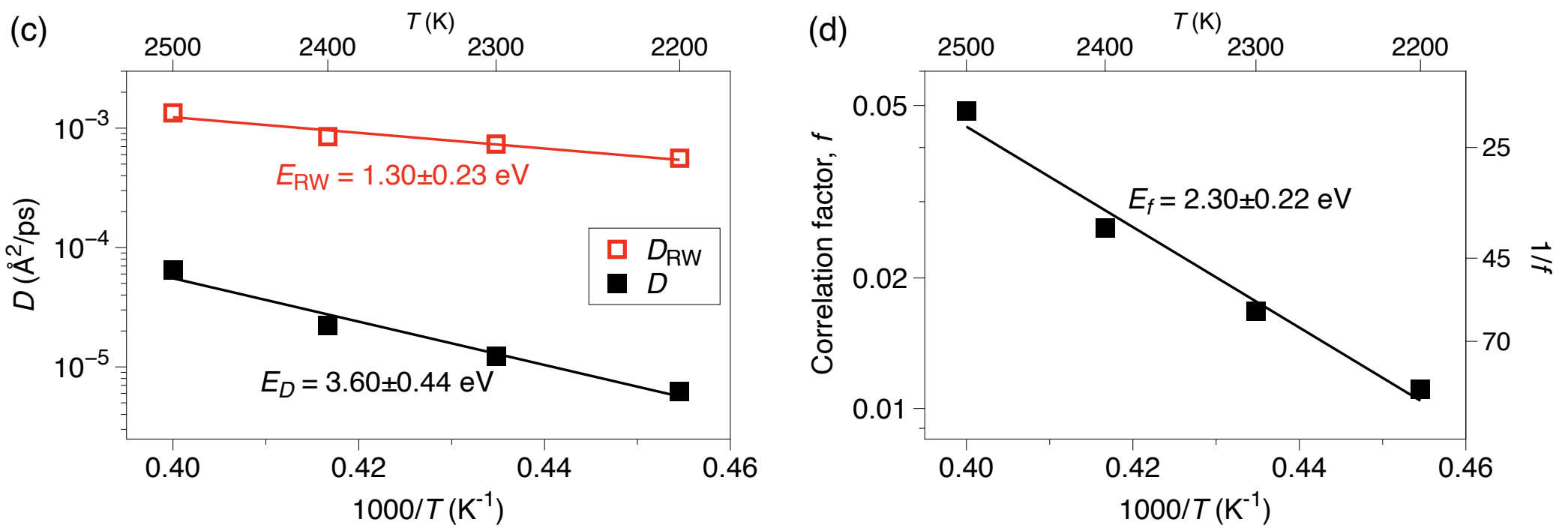

Lennard-Jones (LJ) (single component) glass

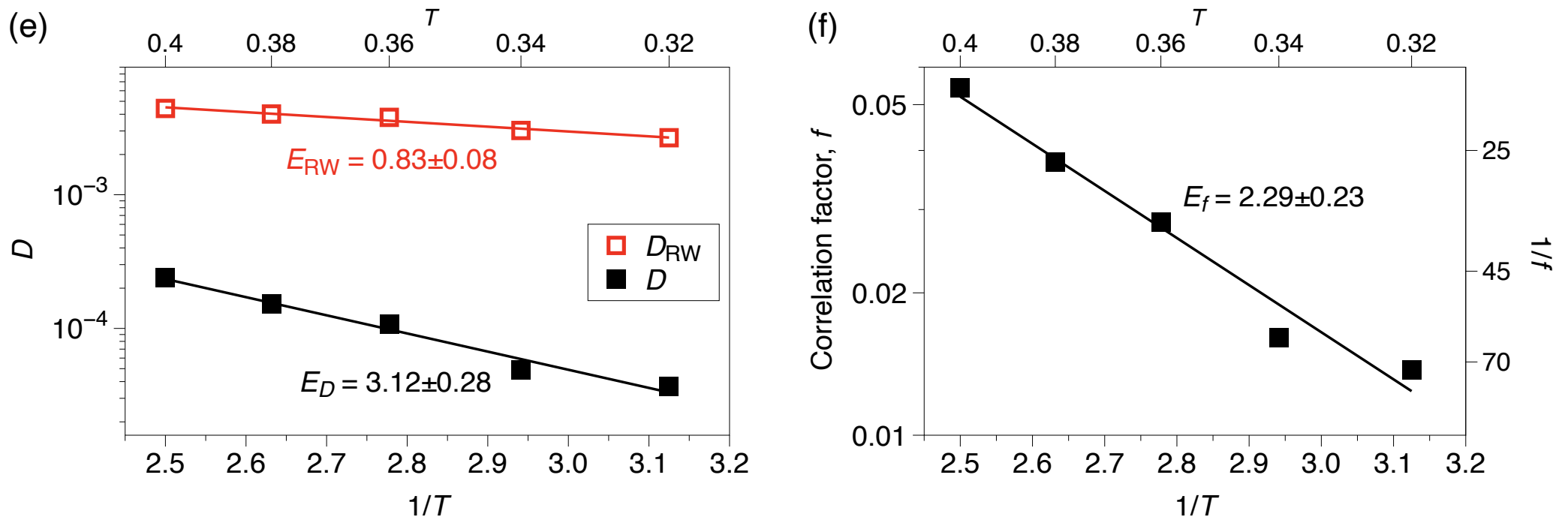

**Fig. S6 |** Arrhenius plots of the total diffusion coefficient $D$ and the random-walk diffusion coefficient $D_{RW}$ in three glassy systems: (a) $Ni_{80}P_{20}$, (c) $SiO_2$, and (e) LJ glass. The correlation factor $f$ for the three systems is shown in (b), (d), and (f), respectively. For analyzing the random-walk diffusion in $Ni_{80}P_{20}$ and $SiO_2$, a displacement threshold distance of 1 Å and a time interval of 1 ps was used. In LJ system, a displacement threshold distance of 0.4 and a time interval of 1000 steps (with timestep, $\Delta t = 0.002$) was used. The LJ units are dimensionless.

S6. Variation of diffusion coefficients and correlation factor ($f$) with cooling-rate:

In Fig. S7, we show the temperature dependence of the MD diffusion coefficient ($D$), the random-walk diffusion coefficient ($D_{RW}$) constructed from MD trajectories using a displacement threshold of 1 Å, and the correlation factor ($f$) for $Cu_{50}Zr_{50}$ cooled at different rates over the temperature range of 640 – 700 K. The activation barriers are listed in Table S1.

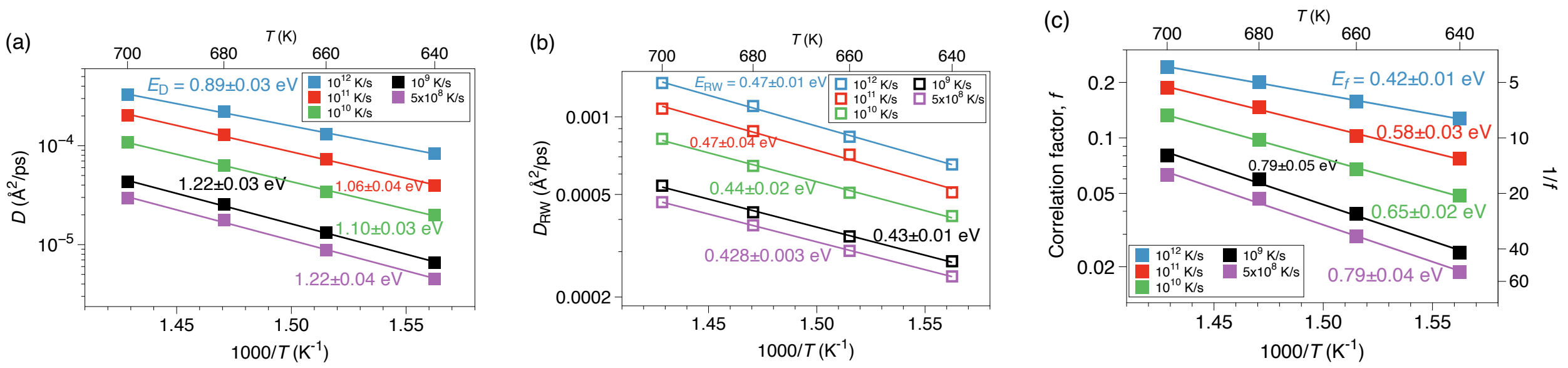

**Fig. S7 |** Arrhenius plots of (a) the MD diffusion coefficient $D$, (b) the random-walk diffusion coefficient $D_{RW}$, and (c) the correlation factor $f$ for $Cu_{50}Zr_{50}$ cooled at different rates.

Table S1. Cooling-rate dependence of activation barriers associated with diffusion coefficient, random-walk diffusion coefficient, and correlation factor for $Cu_{50}Zr_{50}$ in the glassy state.

| Cooling rate (K/s) | $E_D$ (eV) | $E_{RW}$ (eV) | $E_f$ (eV) |
|---|---|---|---|
| $5 \times 10^8$ | 1.22 | 0.43 | 0.79 |
| $1.66 \times 10^9$ | 1.22 | 0.43 | 0.79 |
| $10^{10}$ | 1.10 | 0.44 | 0.65 |
| $10^{11}$ | 1.06 | 0.47 | 0.58 |
| $10^{12}$ | 0.89 | 0.47 | 0.42 |

S7. Estimation of local rearrangement barriers from iso-configurational ensemble MD simulations and Activation-Relaxation Technique (ART):

An independent approach to demonstrate that the rearrangement barriers overcome by atoms in MD simulations are generally much smaller than the overall activation energy for diffusion $E_D$ is through iso-configurational ensemble runs on glass configurations. Briefly, the fraction of iso-configurational trajectories in which an atom undergoes a rearrangement, called the rearrangement probability ($P_R$), provides an estimate of its rearrangement barrier $E_{act}$. Intuitively, a larger $P_R$ corresponds to a smaller $E_{act}$, and vice versa. Here, we consider an atom to have rearranged if its inherent-structure displacement from the initial position reaches at least 1 Å (i.e., $d$ = 1 Å). To convert these rearrangement probabilities into barriers, an estimate of the attempt frequency ($\nu$) relevant to MD simulations is required; in Section S8, we show that $\nu = 10^{12}$ s$^{-1}$ provides a good estimate and also give details of the relation connecting $P_R$ and $E_{act}$. Using MD configurations obtained at 640 K as the starting point for the iso-configurational runs, roughly 80% of atoms undergo at least one rearrangement across the ensemble (i.e., $P_R > 0$). The resulting distribution of activation barriers is shown in Fig. S8*a*, again illustrating that the vast majority of atoms have rearrangement barriers smaller than $E_D$.

Lastly, we employ the Activation-Relaxation Technique (ART) to characterize rearrangement barriers through exploration of the inherent-structure energy landscape. A brief description of ART is provided in the Methods section. Fig. S8*b* shows the resulting distribution of activation barriers for inherent structures obtained at 640 K. Although the distribution peaks near ~1.4 eV, it also reveals a substantial population of lower-barrier rearrangements, which are preferentially sampled during MD simulations. Note that it is difficult to quantitatively compare the iso-configurational distribution and that from ART. The iso-configurational distribution is only from atoms that are displaced during thermally excited rearrangement events and therefore mostly include low barrier events. However, ART is all the barriers found by the algorithm for rearrangement events and therefore might include contributions from both low and high barrier events for a given atom. Despite these differences, both show many low-barrier events.

Taken together, the NEB analysis (from the main text) and the iso-configurational and ART analyses (described here) demonstrate that MD-accessible rearrangements typically involve barriers far smaller than $E_D$. This indicates that the large diffusion activation energy does not arise from correspondingly high local barriers but instead reflects the cumulative effect of correlated hops and back-and-forth motion that suppress net transport.

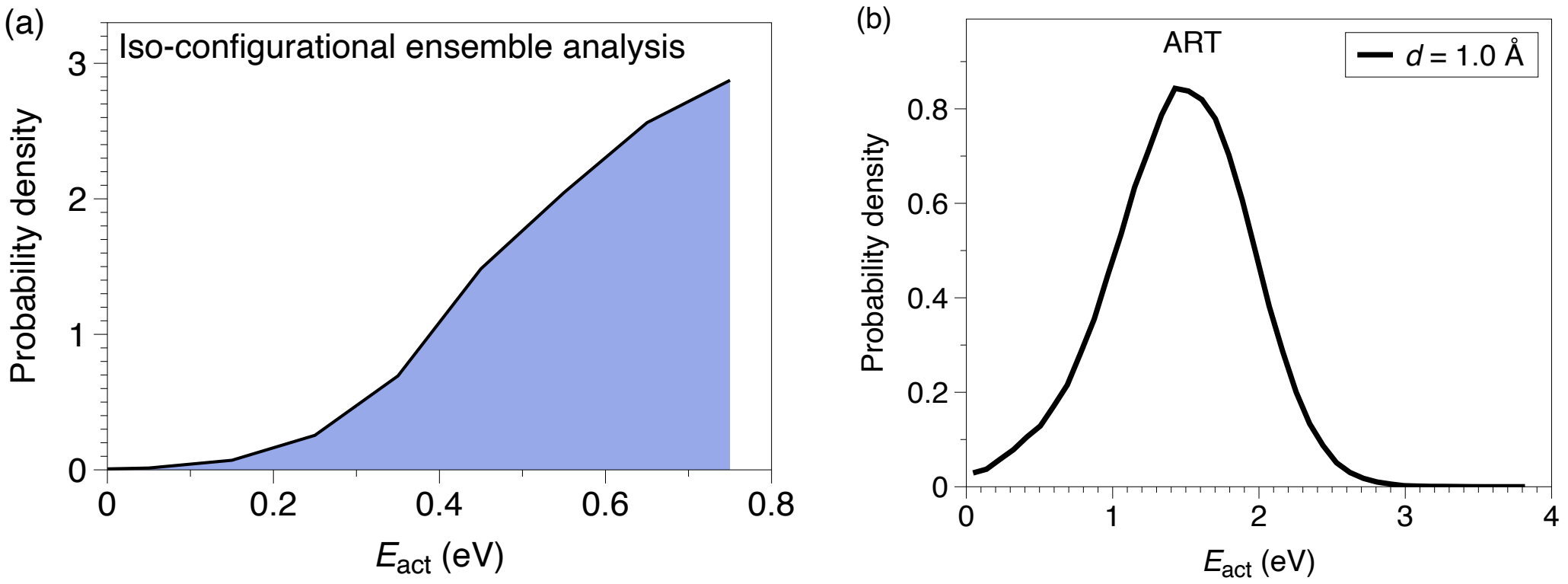

**Fig. S8 |** (a) Distribution of activation energies obtained from iso-configurational ensemble rearrangement probability values. (b) Activation-barrier distribution obtained using ART applied to inherent structures of MD glass configurations.

S8. Going from the probability of rearrangement ($P_R$) obtained from iso-configurational ensemble runs to the rearrangement barrier ($E_{act}$) of atoms:

Starting from 10 independent atomic configurations obtained at 640 K, we carried out 1,000 iso-configurational ensemble runs (each of 2 ps duration) for each configuration at both 640 K and 1200 K. The rationale for performing iso-configurational runs at 1200 K on configurations obtained at 640 K is discussed below. The rearrangement probability $P_R$ of an atom is defined as the probability that it undergoes a rearrangement relative to its initial position during an iso-configurational run. An atom is considered to have rearranged if, within the inherent structures extracted from the 2 ps MD run, its displacement from the initial position reaches at least 1 Å (i.e., $d = 1$ Å). Atomic positions were monitored every 0.25 ps, allowing us to detect rearrangements even if the atom subsequently returned to its original position within the 2 ps window. Figs. S9a and S9b shows the distribution of $P_R$ for all atoms that rearranged at least once across the 1,000 iso-configurational runs at 640 K and 1200 K, respectively.

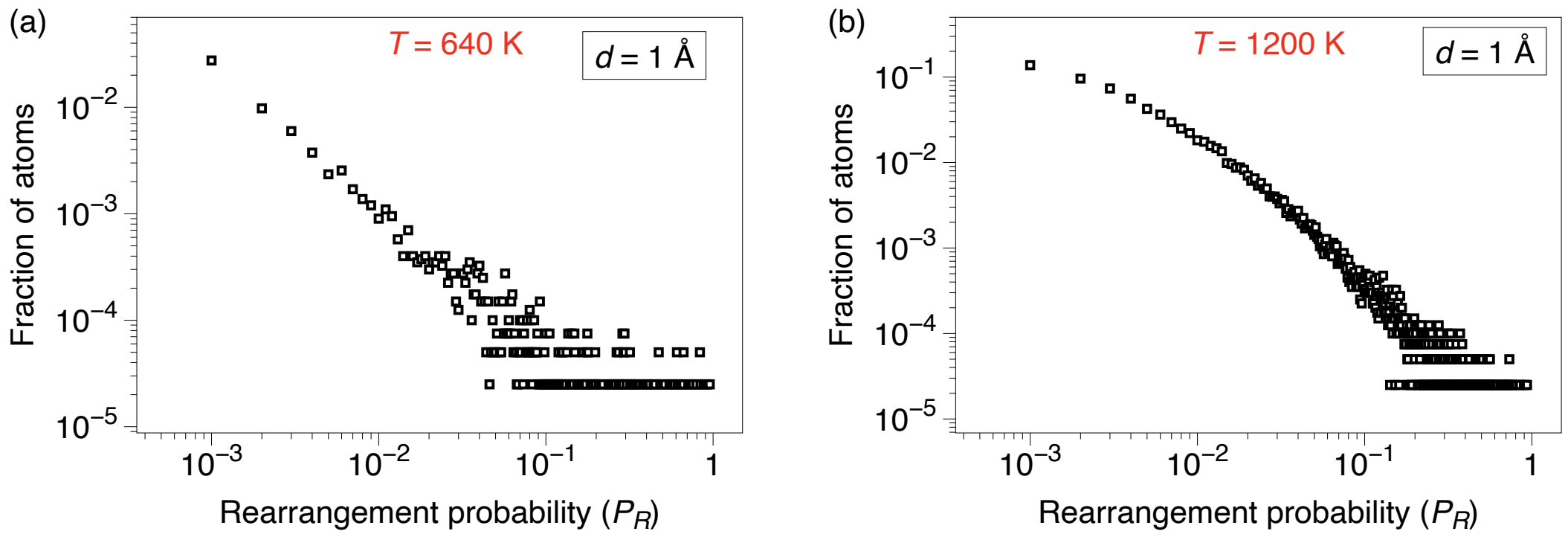

**Fig. S9 |** Distribution of the rearrangement probability $P_R$ for the rearranging atoms at (*a*) 640 K and (*b*) 1200 K. At 640 K, ~8% of atoms (3,059 out of 40,000) exhibit a non-zero $P_R$, whereas at 1200 K this fraction increases to more than 80% (32,514 out of 40,000).

Transition state theory (TST) provides a simple connection between the activation energy $E_{act}$ and $P_R$. According to TST, the rate constant for an atom's rearrangement over a barrier $E_{act}$ is given by

$$k = \nu \cdot \exp(-E_{act}/k_B T), \tag{S1}$$

where $\nu$ is the attempt frequency (typically $10^{12} – 10^{13}$ $s^{-1}$), $k_B$ is the Boltzmann constant, and $T$ the absolute temperature. Assuming that barrier crossing follows a Poisson process with constant rate, the probability that the first crossing occurs within time $t$ is

$$P_R = 1 - \exp(-kt). \tag{S2}$$

Combining Eqs. (S1) and (S2), we obtain

$$P_R = 1 - \exp(-\nu t \exp(-E_{act}/k_B T))$$

$$\Rightarrow E_{act} = k_B T . \ln\left(-\frac{\nu t}{\ln(1-P_R)}\right). \tag{S3}$$

Eq. (S3) allows us to estimate the rearrangement barrier $E_{act}$ directly from the measured $P_R$ (3). A smaller $P_R$ corresponds to a larger $E_{act}$, and vice versa. With 1000 iso-configurational runs, the lowest possible (non-zero) value of $P_R$ is 0.001, and this limits the maximum value of $E_{act}$ that can be assigned to atoms. However, for a fixed value of $P_R$, the inferred $E_{act}$ increases linearly with temperature. This provides a practical and efficient way to access higher activation barriers: performing the iso-configurational runs at an elevated temperature increases the measurable range of $P_R$ without requiring an impractically large number of iso-configurational runs.

We now explain the rationale for how we chose to perform high-temperature iso-configurational runs at 1200 K. As shown in Fig. S9a, at 640 K, only ~8% of the atoms exhibit a nonzero $P_R$ within a 2 ps window. Performing the iso-configurational runs at a higher temperature increases the fraction of atoms that undergo rearrangements, allowing more comprehensive estimation of the barrier distribution. Although the relative contributions of low- and high-barrier events differ at high versus low temperatures and dynamics cannot be directly extrapolated, our runs are sufficiently short (2 ps) such that the aim is merely to allow atoms to overcome any barrier present in the inherent structures, rather than to replicate long-time dynamics at 640 K. This procedure thus enables a more complete estimation of activation barriers for the system.

A natural question is how high a temperature we can use for the iso-configurational runs while still capturing the same set of activation barriers that govern rearrangements at 640 K. This can be assessed by relating the rearrangement probabilities $P_{R,1}$ and $P_{R,2}$ at temperatures $T_1$ and $T_2$ through Eq. (S3). Setting the activation energies equal at the two temperatures gives

$$k_B T_1 . \ln\left(-\frac{\nu t}{\ln(1-P_{R,1})}\right) = k_B T_2 . \ln\left(-\frac{\nu t}{\ln(1-P_{R,2})}\right). \quad \text{(S4)}$$

The value of the attempt frequency $\nu$ is a free parameter in Eq. (S4) and should be chosen to produce consistency with the observed data.

Fig. S10a compares the number of rearrangements at 640 K and 1200 K for atoms that rearranged at least once at both the temperatures. The red line shows the prediction from Eq. (S4) using $\nu = 10^{12}$ s$^{-1}$, a physically reasonable choice. Although the data exhibit substantial scatter, the overall trend follows the theoretical expectation, particularly for atoms with small $P_R$. Fig. S10b shows a moving-average representation of the same data, making the underlying agreement more apparent. For comparison, we also show in Fig. S10b the varation of Eq. (S4) with $\nu = 5 \times 10^{11}$ s$^{-1}$ and $\nu = 5 \times 10^{12}$ s$^{-1}$. In contrast, Figs. S10c and S10d show the corresponding comparison when 1500 K is used for the high-temperature simulations. In this case, the data display large scatter across the entire range of $P_R$, and no meaningful alignment with the TST prediction is observed. This indicates that 1500 K accesses a qualitatively different set of barriers and therefore cannot reliably capture the rearrangement barriers relevant at 640 K.

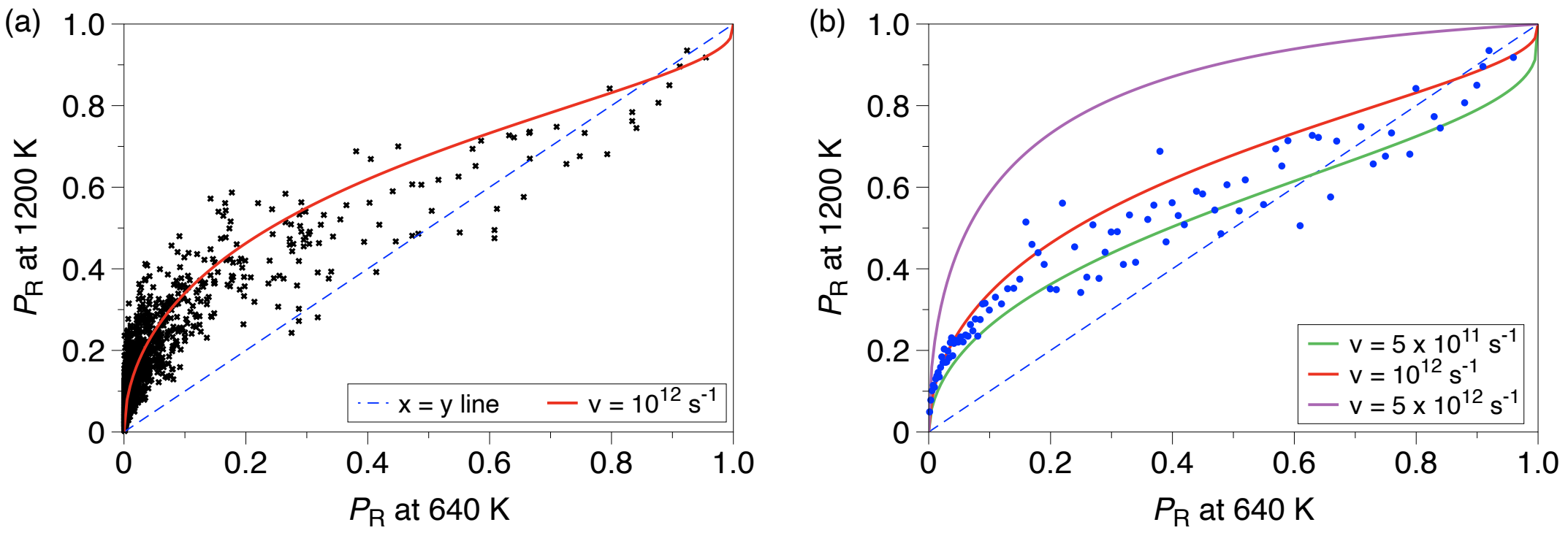

—

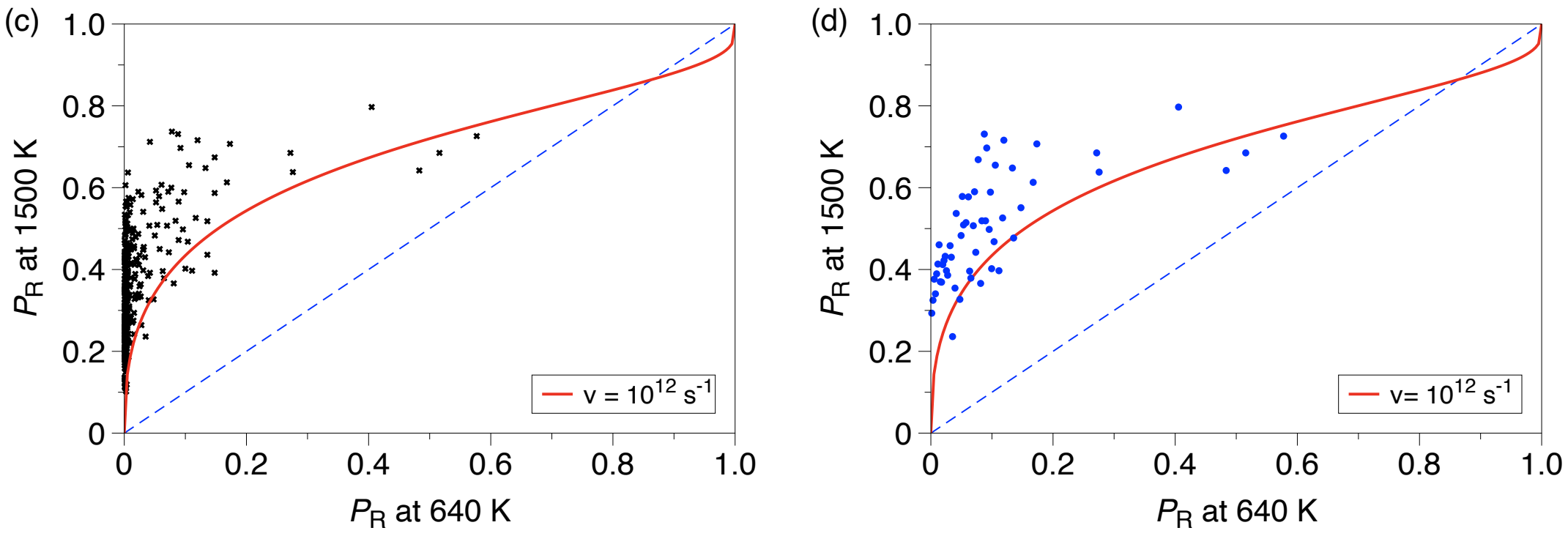

**Fig. S10 |** (a) Comparison of the rearrangement probability $P_R$ for the rearranging atoms at 640 K versus 1200 K. In (b), we show a moving-average representation of the data in (a). (c) $P_R$ comparison for the rearranging atoms at both 640 K and 1500 K. (d) Moving-average representation of the data in (c).

## S9. Time evolution of mean-squared displacement (MSD) of surface atoms:

In $Cu_{50}Zr_{50}$, we find that the surface diffusion does not reach a fully diffusive regime within the accessible simulation times. Fig. S11a shows the time evolution of the two-dimensional mean-squared displacement (MSD) of surface atoms at 700 K, obtained from direct MD simulations and from the reconstructed random-walk trajectories. The dotted line indicates a unit slope in the log-log representation, corresponding to ideal Fickian diffusion. Both MSDs exhibit persistent non-Fickian behavior even at the longest times examined, extending to mean-squared displacements exceeding 100 Å$^2$. The MD MSD displays a subdiffusive log-log slope of approximately 0.7, whereas the corresponding random-walk MSD exhibits a substantially larger slope of ~0.9. Fig. S11b shows the convergence behavior of the surface correlation factor, $f_s$ during our MD simulations. In Fig. S11c, we compare the surface and bulk diffusion coefficients at different temperatures below $T_g$.

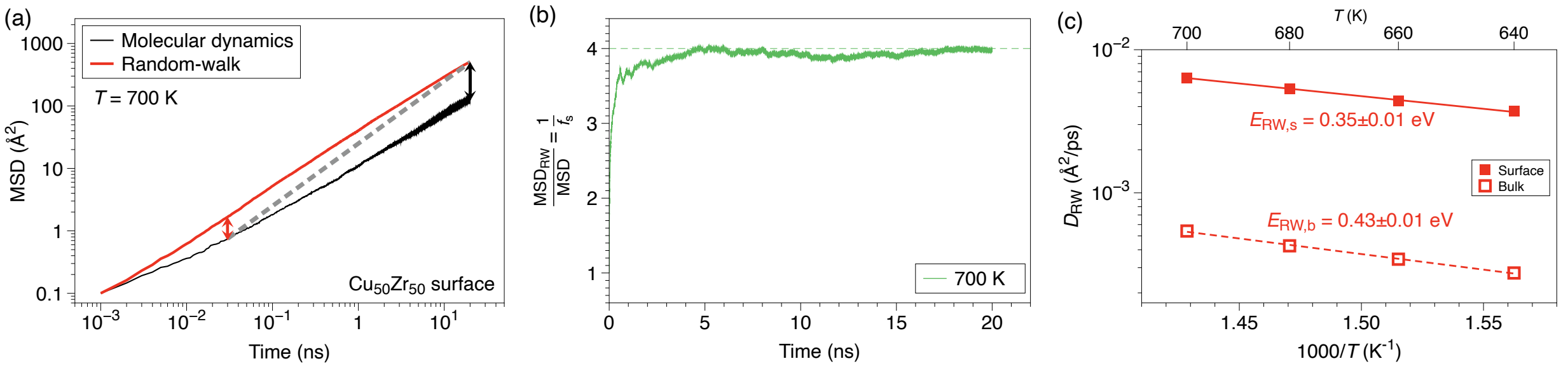

**Fig. S11 |** (a) Time evolution of the MSD of surface atoms obtained from direct MD simulations and from reconstructed random-walk trajectories, shown on a log–log scale. The dotted line indicates a unit slope corresponding to ideal Fickian diffusion. The black and red double-headed arrows highlight the extent of the non-Fickian diffusion for the MD MSD and the random-walk MSD, respectively. (b) Convergence of the surface correlation factor $f_s$. Shown is the time evolution of the ratio between the random-walk MSD and the MD MSD for surface atoms, which by definition equals $1/f_s$. The ratio approaches a constant value at long times, demonstrating convergence of the correlation factor despite the non-Fickian nature of both the MD and random-walk MSDs. (c) Arrhenius plots of surface and bulk diffusion coefficients in the glassy state.

## S10. Nature of aging from molecular dynamics simulations:

In this work, to study the dynamics at a given target temperature, we start from the final configuration obtained after quenching to 300 K and anneal it at the target temperature for 1 ns in the NPT ensemble. Subsequently, property calculations are performed in the NVT ensemble. As shown in Fig. S12, annealing from 300 K to the temperature of interest introduces minimal aging effects (potential energy is quite stable within our simulation time) when compared to configurations obtained directly during the quenching process at the same temperature.

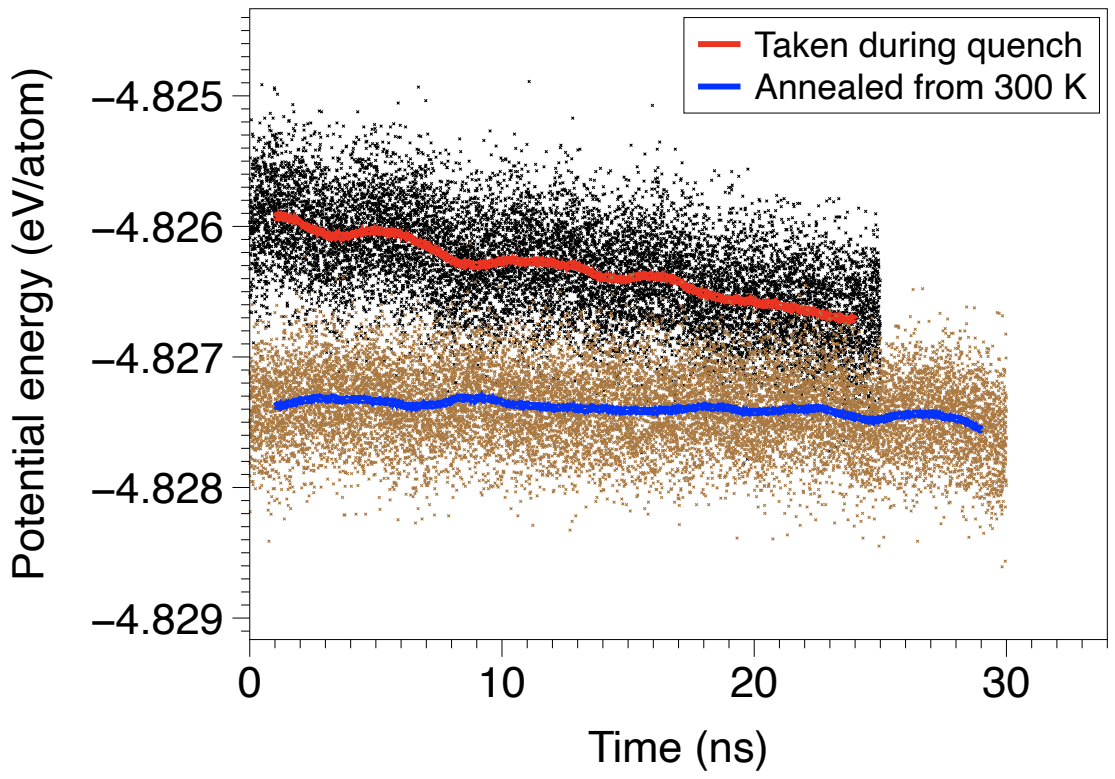

**Fig. S12 |** Comparison of inherent potential energy obtained using the two preparation protocols described in the text at 660 K in $Cu_{50}Zr_{50}$. Symbols denote the instantaneous inherent-structure energies of the MD configurations, while the solid lines show the corresponding moving averages.

S11. Pre-exponential factors: Below we summarize pre-factors for all the exponential fits used in this work.

| Quantity | Description | Pre-exponential factor |
|---|---|---|
| $D$ | Fig. 1a ($Cu_{50}Zr_{50}$) | 26602 Å$^2$/ps |
| $D_{RW}$ | Fig. 1a | 0.686 Å$^2$/ps |
| $f$ | Fig. 1b | 38753 |
| $D_s$ | Fig. 6a | 739 Å$^2$/ps |
| $D_{s,RW}$ | Fig. 6a | 2.14 Å$^2$/ps |
| $f_s$ | Fig. 6b | 346 |
| $D$ | Fig. S3b (crystalline nickel) | 1.1937 Å$^2$/ps |
| $D_{RW}$ | Fig. S3b | 1.5232 Å$^2$/ps |
| $D_{RW}$ | Fig. S4a | 0.3749 Å$^2$/ps (0.25 Å)<br>0.3962 Å$^2$/ps (0.5 Å)<br>0.6864 Å$^2$/ps (1.0 Å)<br>2.1616 Å$^2$/ps (1.5 Å)<br>32.2343Å$^2$/ps (2.0 Å) |
| $f$ | Fig. S4b | 70940 (0.25 Å)<br>67121 (0.5 Å)<br>38753 (1.0 Å)<br>12305 (1.5 Å)<br>824 (2.0 Å) |
| $D$ | Fig. S6a ($Ni_{80}P_{20}$) | 8657 Å$^2$/ps |
| $D_{RW}$ | Fig. S6a | 6.8702 Å$^2$/ps |
| $f$ | Fig. S6b | 1260 |
| $D$ | Fig. S6c ($SiO_2$) | 1009.23 Å$^2$/ps |
| $D_{RW}$ | Fig. S6c | 0.52 Å$^2$/ps |
| $f$ | Fig. S6d | 1940.69 |
| $D$ | Fig. S6e (LJ) | 0.580 |
| $D_{RW}$ | Fig. S6e | 0.036 |
| $f$ | Fig. S6f | 16.074 |
| $D$ | Fig. S7*a* | 19320 Å$^2$/ps (5x10$^8$ K/s)<br>26602 Å$^2$/ps (10$^9$ K/s)<br>8667 Å$^2$/ps (10$^{10}$ K/s)<br>8616 Å$^2$/ps (10$^{11}$ K/s)<br>920 Å$^2$/ps (10$^{12}$ K/s) |
| $D_{RW}$ | Fig. S7b | 0.56 Å$^2$/ps (5x10$^8$ K/s)<br>0.6864 Å$^2$/ps (10$^9$ K/s)<br>1.2925 Å$^2$/ps (10$^{10}$ K/s)<br>2.8752 Å$^2$/ps (10$^{11}$ K/s)<br>3.5610 Å$^2$/ps (10$^{12}$ K/s) |
| $f$ | Fig. S7c | 34280 (5x10$^8$ K/s)<br>38753 (10$^9$ K/s)<br>6705 (10$^{10}$ K/s)<br>2996 (10$^{11}$ K/s)<br>258.50 (10$^{12}$ K/s) |
| $D_{RW,s}$ | Fig. S11c | 2.137 Å$^2$/ps |
| $D_{RW,b}$ | Fig. S11c | 0.686 Å$^2$/ps |

References:

1. G. Bonny, D. Terentyev, R. C. Pasianot, S. Poncé, A. Bakaev, Interatomic potential to study plasticity in stainless steels: the FeNiCr model alloy. *Model. Simul. Mat. Sci. Eng.* **19**, 085008 (2011).
2. K. Compaan, Y. Haven, Correlation factors for diffusion in solids. *Transactions of the Faraday Society* **52**, 786–801 (1956).
3. A. Annamareddy, B. Wang, P. M. Voyles, D. Morgan, Distribution of atomic rearrangement vectors in a metallic glass. *J. Appl. Phys.* **132**, 195103 (2022).